\documentclass[journal,twoside,web]{ieeecolor}
\usepackage{generic}
\usepackage{cite}
\usepackage{amsmath,amssymb,amsfonts}
\usepackage{algorithmic}
\usepackage{graphicx}
\usepackage{textcomp}
\usepackage{subfigure}
\allowdisplaybreaks[1]

\def\BibTeX{{\rm B\kern-.05em{\sc i\kern-.025em b}\kern-.08em
    T\kern-.1667em\lower.7ex\hbox{E}\kern-.125emX}}
\begin{document}
\title{Approximation-free control based on the bioinspired reference model for suspension systems with uncertainty and unknown nonlinearity}
\author{Xiaoyan Hu, Guilin Wen, Shan Yin, Zhao Tan, and Zebang Pan
\thanks{Xiaoyan Hu, Shan Yin, Zhao Tan and Zebang Pan are with State Key Laboratory of Advanced Design and Manufacture for Vehicle Body, Hunan University, Changsha, Hunan 410082, China.(e-mail:nicole0106@hnu.edu.cn; shan\_yin@hnu.edu.cn; tanzhao@hnu.edu.cn; zbpan@hnu.edu.cn)}
\thanks{Guilin Wen is with: 1 State Key Laboratory of Advanced Design and Manufacture for Vehicle Body, Hunan University, Changsha, Hunan 410082, China. 2 School of Mechanical Engineering, Yanshan University, Qinhuangdao, Hebei 066004, China(e-mail: glwen@hnu.edu.cn)}
}

\maketitle

\begin{abstract}
Uncertainty and unknown nonlinearity are often inevitable in the suspension systems, which were often solved using fuzzy logic system (FLS) or neural networks (NNs). However, these methods are restricted by the structural complexity of the controller and the huge computing cost. Meanwhile, the estimation error of such approximators is affected by adopted adaptive laws and learning gains. Thus, in view of the above problem, this paper proposes the approximation-free control based on the bioinspired reference model for a class of uncertain suspension systems with unknown nonlinearity. The proposed method integrates the superior vibration suppression of the bioinspired reference model and the structural advantage of the prescribed performance function (PPF) in approximation-free control. Then, the vibration suppression performance is improved, the calculation burden is relieved, and the transient performance is improved, which is analyzed theoretically in this paper. Finally, the simulation results validate the approach, and the comparisons show the advantages of the proposed control method in terms of good vibration suppression, fast convergence, and less calculation burden.
\end{abstract}


\section{Introduction}
\label{sec:introduction}
\IEEEPARstart{T}{he} suspension system is one of the important components in vehicles to ensure passengers’ ride comfort and safety by isolating the vibration from road excitation and adapting to the complexity of tough roads. Generally, there are three types of suspension systems: passive suspension, semi-active suspension, and active suspension. Passive suspensions absorb vibration mainly through dampers and springs installed inside, passively.The contradictory ride comfort requirements and handling stability cannot be satisfied simultaneously due to the fixed system parameters. Thus, the semi-active suspension is proposed to improve adaptation by adjusting the damper and spring force, but it cannot be controlled optimally according to external input. Compared to passive and semi-active suspension, active suspensions isolate vibration and offer more comfort effectively and flexibly by producing an active force, whose design and analysis get the most attention from researchers. Due to the simple structure, linear active suspension models have been studied for decades. Some feedback control schemes, such as feedback control \cite{b1,b2,b3,b4} and $H_\infty$ control method with LMI (Linear matrix inequality)\cite{b2,b3,b5,b6}, were implemented smoothly to achieve significant results in these models. However, most of the existing results focused on linear active suspension systems with delay or uncertainty in recent works. Besides, the main form of control methods is feedback gain structure essentially. Furthermore, some linear control methods cannot be further applied in nonlinear systems. Thus, many researchers are investigating nonlinear active suspension system, which has more realistic value and engineering applicability. Various synthetical control methods are widely studied to address the nonlinearity in active suspension systems. In Ref. \cite{b7}, considering the highly complex nonlinearity in the hydraulic actuator, the higher-order terminal sliding mode control was established to stabilize uncertain suspension systems. Adaptive backstepping control was applied to uncertain nonlinear suspension systems considering input delay \cite{b8} and hard constraints \cite{b9}. The adaptive control was established in Ref. \cite{b10} with synchronization control to synchronize the height of four suspensions of the vehicle. 

 Owing to the diversity of passengers, measurement technique limitation, and inherent dynamic property, uncertainty and unknown nonlinearity are inevitable in suspension systems. In order to tackle these issues, approximation methods, such as neural networks (NNs) and fuzzy logic systems (FLS) are extensively investigated by cooperating different control strategies. In Ref. \cite{b11}, with the assistance of NNs approximating uncertain dynamics, the adaptive finite-time control was proposed. The stability and ride comfort of the half suspension system could be achieved during a finite time. For a class of uncertain suspension systems with the time-varying vertical displacement and speed constraints, the adaptive controller based on the NNs approximator was established in Ref. \cite{b12}. Considering the limitation of communication resources, the event-trigger controller with NNs approximating the unknown term was adopted to release the communication burden on electromagnetic actuator to the controller \cite{b13}. The approximator FLS generally cooperates with different control methods including adaptive backstepping control and sliding mode control for a class of suspension systems with nonlinearity or uncertainty. In Ref. \cite{b14}, the suspension system's input delay and unknown nonlinearity were considered and handled by compensation scheme and FLS, respectively. Then the adaptive finite-time fuzzy control scheme was proposed. In Ref. \cite{b15}, fuzzy logic control was designed by combining PID control to reduce the vibration. The comparison simulation verified its good performance. Shalabi et al \cite{b16} utilized the Neuro-fuzzy inference system to air suspension system and guaranteed ride comfort and handling stability by controlling solenoid valves. An experiment was carried out to verify the effectiveness of the controller. In Ref. \cite{b17,b18}, the sliding mode control was integrated with FLS to enhance ride comfort. Specifically, the NNs and FLS were combined in Ref. \cite{b17}. Both NNs and FLS are powerful tools to deal with the unknown nonlinearity and uncertainty. However, with the increase in the number of neurons in NNs, the computational burden increases, although the approximation error gets smaller. This is also observed with the FLS. The approximating results become precise with the increase in logic rules, but the calculation becomes complicated.

 Bechlioulis et al. \cite{b19} first proposed the prescribed performance control (PPC) to simplify the design procedure and relieve the calculation burden,. The main idea of PPC is to preset a performance function that defines the controlled plant’s convergence rate, overshoot range and signal convergence boundary. Thus, it can ensure both transient and steady state performance of systems. Then they extended the method to the MIMO system \cite{b20} and combined it with the partial state feedback method for an unknown nonlinear system \cite{b21}. Moreover, its structure was further simplified \cite{b22}, similar to the recursion scheme of backstepping control but avoided the problem of ‘explosion of complexity’ existing in backstepping. Subsequently, this method was applied to nonlinear suspension systems \cite{b23,b24,b241} and some other systems\cite{b21,b25,b26,b27,b28} like servo mechanisms and MEMS gyroscopes and obtained considerable results. Unlike the control scheme using approximation function, such as NNs and FLS, the PPC does not require any other control methods’ assistance to complete the control goal, and its simple scheme reduces computation time meanwhile. However, the simple structure also results in its sensibility to rapidly changing signals and limits the ability of PPC to suppress vibration for suspensions. Thus, there is a need to find an auxiliary method to enhance its vibration suppression performance.
Although the performance of active suspension systems is improved a lot by applying various control methods like that mentioned above, vibration suppressing is still an important issue to be further enhanced due to the complex dynamic structure and changeable working environment. Pan et.al \cite{b29} first used the bioinspired structure as a reference model in a nonlinear suspension system to isolate more vibration, while this structure was initially regarded as a vibration isolator in Ref. \cite{b30}. Subsequently, various control methods such as fuzzy adaptive control, fuzzy sliding mode control and adaptive neural network control were implemented to enhance the performance of the suspension system with the bioinspired reference model\cite{b31,b32,b33}, considering uncertainty and unknown nonlinearity. The main merit of introducing the bioinspired model is to utilize the beneficial nonlinearity inside, which is useful to suppress vibration of the system. However, it should be noted that the fuzzy method and neural networks were the primary tools to deal with uncertainty and unknown nonlinearity in these works. Their common defects of complex structure and computing redundancy as mentioned before remain still. Therefore, this paper proposes the approximation-free control based on the bioinspired reference model for the suspension system with uncertainty and unknown nonlinearity. 

The primary contributions of this paper are concluded as follows: Firstly, compared to controllers with the FLS or NNs approximators in \cite{b31,b32,b33}, the proposed control scheme is a simple structure that has a similar but more concise recursive framework of backstepping control and possesses the merit to handle inexact model without approximation. The simple structure releases the calculation burden and saves significant time. Furthermore, the acceleration decreases due to the bioinspired reference model and PPF, resulting in improved comfort but guaranteed road-handling. Secondly, due to the prescribed performance function in the proposed controller, the tracking error can be converged sooner and the ultimate convergence bound can be remained in a smaller region than the comparison subject (fuzzy adaptive control (FAC)\cite{b31}). Then it enhances the system convergence performance and helps the active suspension system respond quickly and precisely in reality. In particular, for the first time in this paper, the superior convergence performance of PPF is theoretically compared to FLS. This helps in the practical design of the controller.

The rest of this paper is organized as follows. In Section 2, the suspension system model and the bioinspired model are formed and some prior conditions are presented. The scheme of approximation-free control based on the bioinspired reference model is established and the stability proof and convergence performance analysis are given in Section 3. Then, the validity and superiority are verified by simulation and comparison in Section 4. Finally, the conclusion is presented in Section 5.

\section{Problem formulation}
The whole structure of the suspension system tracking the bioinspired model under the approximation-free control can be described in Fig 1. For the part of the active suspension system in Fig 1, we can get the following dynamic equations of the suspension system based on Newton's second law.

\begin{equation}\label{eq1}\begin{array}{l}
		{m_s}{{\ddot z}_s} + {F_d}({{\dot z}_s},{{\dot z}_u}) + {F_s}({z_s},{z_u}) = u\\
		{m_u}{{\ddot z}_u} - {F_d}({{\dot z}_s},{{\dot z}_u}) - {F_s}({z_s},{z_u}) + {F_t}({z_u},{z_r}) + \\
		{F_b}({{\dot z}_u},{{\dot z}_r}) =  - u
	\end{array}\end{equation}
where 
$$\begin{array}{l}
	{F_t} = {k_t}({z_u} - {z_r})\\
	{F_b} = {c_t}({{\dot z}_u} - {{\dot z}_r})
\end{array}$$
 $m_s$ represents sprung mass, which is an uncertain term owing to the change in passengers;  $m_u$ is the unsprung mass; $z_s$  and $z_u$  are absolute displacements of sprung mass, unsprung mass respected to ground, respectively;  $z_r$ denotes the road excitation; $u$ is the input force from the designed controller;  $F_s$ and $F_d$ are unknown nonlinear forces produced by springs and dampers, respectively; Elastic and damping forces of tire are denoted by $F_t$ and $F_b$ , respectively; For the convenience of subsequent analysis, some assumptions for the suspension system should be given here.
 
 $\boldsymbol{Assumption 1}$ The uncertain term in \eqref{eq1} is bounded and assumed to satisfy that ${m_s} \in {\Omega _{{m_s}}} = \left\{ {{m_s}:{m_{s\min }} \le {m_s} \le {m_{s\max }}} \right\} $, where ${m_{s\min }}$ and ${m_{s\max }}$ are lower and upper bound of sprung mass, respectively.
 
 $\boldsymbol{Assumption 2}$ The unknown nonlinear forces $F_s$ and $F_d$ are bounded and continuous, which is described by inequalities:  $\left| {{F_s}} \right| < {\bar F_s}$ and $\left| {{F_d}} \right| < {\bar F_d}$.
 
 $\boldsymbol{Assumption 3}$ The road excitation $z_r$ and corresponding derivative ${\dot z_r}$ are limited, Given the positive constant  ${\bar z_{r1}}$ and ${\bar z_{r2}}$ , they are satisfied that $\left| {{z_r}} \right| < {\bar z_{r1}}$ and $\left| {{{\dot z}_r}} \right| < {\bar z_{r2}}$.
 \begin{figure}[!t]
 	\centerline{\includegraphics[width=\columnwidth]{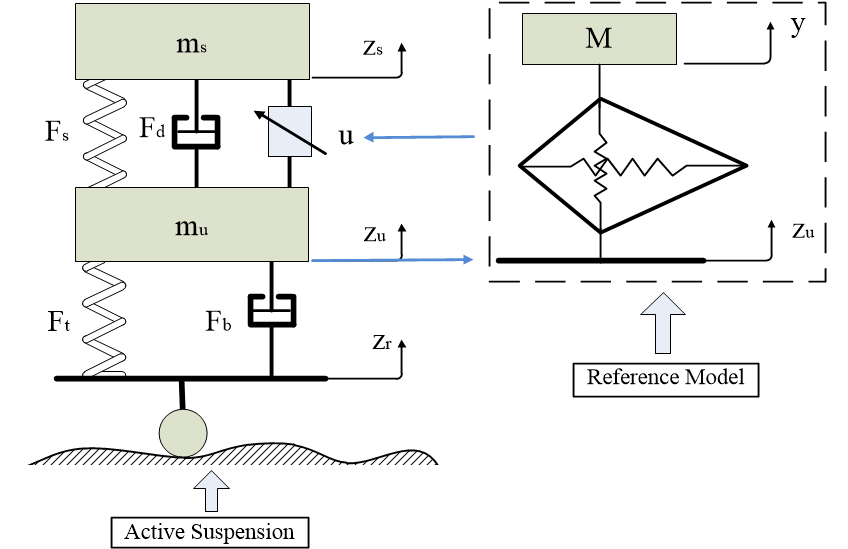}}
 	\caption{Whole structure of suspension system with the bioinspired reference model.}
 	\label{fig_1}
 \end{figure}

We can further rewrite \eqref{eq1} to state-space form. Defining state vector $[{x_1},{x_2},{x_3},{x_4}] = [{z_s},{\dot z_s},{z_u},{\dot z_u}]$, state-space equation can be represented as
\begin{equation}\label{eq2}
	\left\{ \begin{array}{l}
		\begin{split}
			{{\dot x}_1} =& {x_2}\\
			{{\dot x}_2} =& \frac{1}{{{m_s}}}( - {F_d}({x_2},{x_4}) - {F_s}({x_1},{x_3}) + u)\\
			{{\dot x}_3} =& {x_4}\\
			{{\dot x}_4} =& \frac{1}{{{m_u}}}({F_d}({x_2},{x_4}) + {F_s}({x_1},{x_3}) - {F_t}({x_3},{z_r})\\
			&- {F_b}({x_4},{{\dot z}_r}) - u)
		\end{split}	
	\end{array} \right.\end{equation}
Besides, by defining the relative displacement as state, that is, state vector $[{z_1},{z_2}] = [{z_s} - {z_u},{\dot z_s} - {\dot z_u}]$, one can obtain the relative state-space form inferred from \eqref{eq1}
\begin{equation}\label{eq3}\left\{ \begin{array}{l}
	{{\dot z}_1} = {z_2}\\
	{{\dot z}_2} = {{\ddot z}_s} - {{\ddot z}_u} = \theta (\chi ({z_1},{z_2}) + u) - {{\ddot z}_u}
\end{array} \right.
\end{equation}
where the coefficient $\theta  = 1/{m_s}$, and nonlinear term $\chi  =  - {F_s} - {F_d}$. Furthermore, the suspension system should satisfy the following constraints to ensure safety. One is that the dynamic load of the tire during driving is less than the weight of the vehicle so that the tire can keep uninterrupted contact with the road. Another one is that the suspension dynamic deflection in riding can not exceed the max suspension stroke. The following equations express the relationship
\begin{align}
	\begin{split}
\label{eq4}{F_t} + {F_b} <& ({m_s} + {m_u})g
	\end{split}\\
	\begin{split}
\label{eq5}{z_s} - {z_u} <& {z_{\max }}	
\end{split}
\end{align}

\begin{figure}[!t]
	\centerline{\includegraphics[width=\columnwidth]{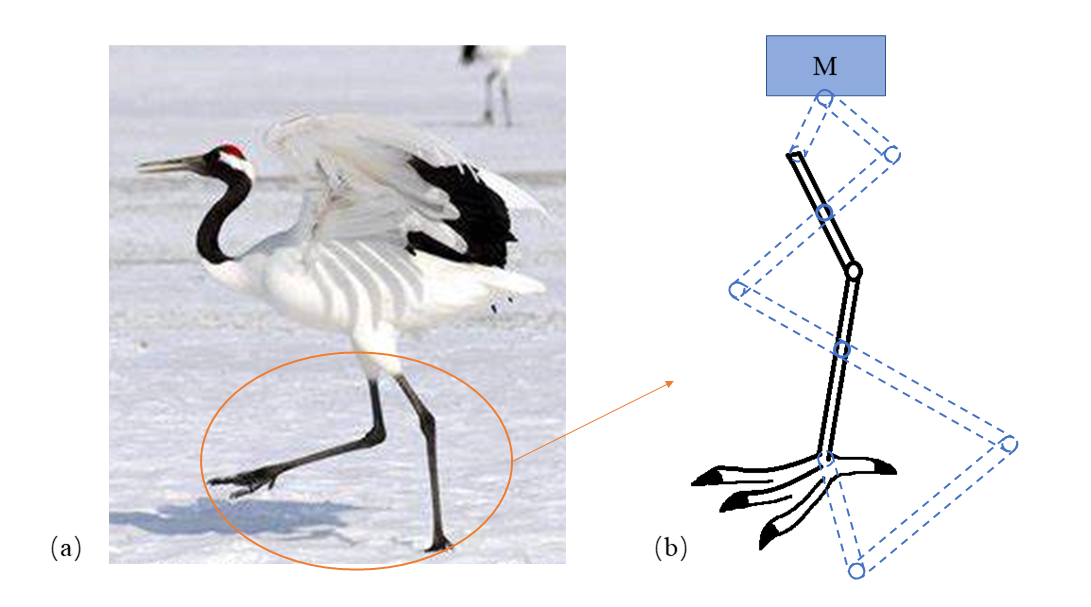}}
	\caption{Dynamic structure of Grus Japonensis’s leg\cite{b30}.}
	\label{fig_2}
\end{figure}

Next, the bioinspired model to be used in control design is discussed. Grus Japonensis's leg inspires the part of the bioinspired reference model in Fig \ref{fig_1}. The Japonensis’s leg structure is displayed in Fig \ref{fig_2} (a) , and the corresponding equivalent dynamic structure is depicted in Fig \ref{fig_2} (b). For a more detailed analysis of bioinspired structure, one can refer to \cite{b30}.

\begin{figure}[!t]
	\centerline{\includegraphics[width=\columnwidth]{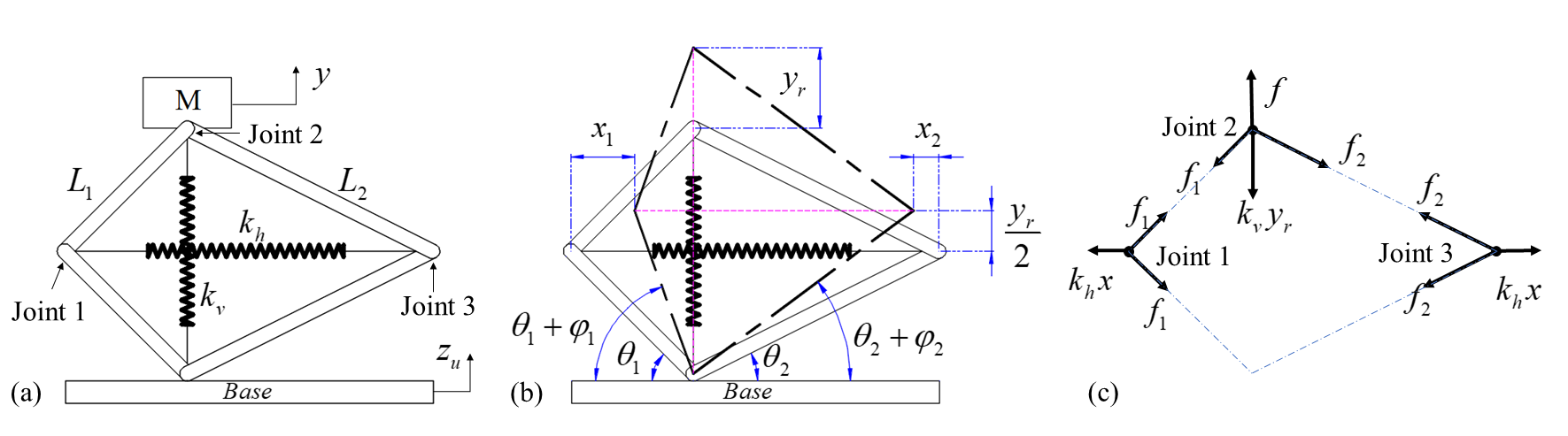}}
	\caption{Geometry structure of the bioinspired model and dynamic analysis.}
	\label{fig_3}
\end{figure}

The bioinspired structure can be further simplified into an X-shape structure, shown in Fig \ref{fig_3} (a), consisting of two shorter rods, two longer rods, one vertical spring, and one horizontal spring. The motion during extension is described in Fig \ref{fig_3} (b). The force condition of different joints is depicted in Fig \ref{fig_3} (c). $M$ is the mass of vibration isolated target.  $x$ and $y$ express the absolute horizontal displacement of joints 1 and 3 and the absolute vertical displacement of $M$, respectively. ${x_1}$ and ${x_2}$ are the relative displacement of joint 1 and joint 3 with respect to the original position, respectively. ${y_r}$ represents the relative displacement of joints 1 and 3 in compression or extension respected to that in the initial position. ${L_1}$ and ${L_2}$ are the length of the rod, which satisfy ${L_1} < {L_2}$. ${\theta _1}$ and ${\theta _2}$ are angles between rod and base, which meet ${\theta _1} > {\theta _2}$. ${\varphi _1}$ and ${\varphi _2}$ are relative angles respected to initial angle when rod rotates. ${k_h}$ and ${k_v}$ represent the stiffness coefficient of horizontal and vertical spring in the model, respectively. ${z_u}$ is the road excitation.

The dynamic analysis of the bioinspired model is given in Appendix \ref{appendix 1}. The nonlinear dynamic equation can be written as follows by defining  ${y_r} = y - {z_u}$. 
\begin{equation}\label{eq6}
M{\ddot y_r} + {h_1}({y_r}) + {k_v}{y_r} + {\mu _1}{\dot y_r} + {\mu _2}{n_x}{h_2}({y_r}){\dot y_r} =  - M{\ddot z_u}
\end{equation}
where ${\mu _1}$ and ${\mu _2}$ are the air damping coefficient and rotational friction coefficient, respectively. ${n_x}$ is the number of joints. ${h_1}({y_r})$ and ${h_2}({y_r})$ can be expressed \eqref{eq7} by defining  ${y_1} = {y_r}$, ${y_2} = {\dot y_r}$, $v({y_1}) = {L_1}\sin {\theta _1} + {{{y_1}} \mathord{\left/{\vphantom {{{y_1}} 2}} \right.\kern-\nulldelimiterspace} 2}$, and ${\sigma _i} = \sqrt {L_i^2 - {v^2}({y_1})} \left(i = 1,2\right)$. Then defining $f({y_1}) = {h_1}({y_1}) + {k_v}{y_1}$, and $g({y_1},{y_2}) = {\mu _1}{y_2} + {\mu _2}{n_x}{h_2}({y_1}){y_2}$, the state-space form of the bioinspired model is obtained, shown in \eqref{eq8}. 

\begin{equation}\label{eq7}
	\left\{\begin{array}{c}
		\begin{split}
			{h_1}({y_1}) =& \frac{{{k_h}}}{2}\left( {{L_1}\cos {\theta _1} + {L_2}\cos {\theta _2} - {\sigma _1}} \right.\left. { - {\sigma _2}} \right)\\
			&\cdot \left( {{{v({y_1})} \mathord{\left/
						{\vphantom {{v({y_1})} {{\sigma _1}}}} \right.
						\kern-\nulldelimiterspace} {{\sigma _1}}} + {{v({y_1})} \mathord{\left/
						{\vphantom {{v({y_1})} {{\sigma _2}}}} \right.
						\kern-\nulldelimiterspace} {{\sigma _2}}}} \right)\\
			{h_2}({y_1}) =& \left( {{1 \mathord{\left/
						{\vphantom {1 {(2{\sigma _1})}}} \right.
						\kern-\nulldelimiterspace} {(2{\sigma _1})}} + {1 \mathord{\left/
						{\vphantom {1 {(2{\sigma _2})}}} \right.
						\kern-\nulldelimiterspace} {(2{\sigma _2})}}} \right)
		\end{split}		
	\end{array} \right.
\end{equation}

\begin{equation}\label{eq8}
\left\{ \begin{array}{l}
	{{\dot y}_1} = {y_2}\\
	{{\dot y}_2} =  - \frac{1}{M}\left( {f({y_1}) + g({y_1},{y_2})} \right) - {{\ddot z}_u}
\end{array} \right.
\end{equation}

Equation \eqref{eq8} represents the relationship of internal parameter and dynamic behavior of the bioinspired structure, which will be used as a reference model for the active suspension system in Fig \ref{fig_1}. It is worth noting that the bioinspired structure containing nonlinearity is beneficial for vibration suppression when designing the controller\cite{b31}.

It is necessary to investigate the stability of the bioinspired model to facilitate subsequent analysis. The conclusion of stability analysis will be given first here.

$\boldsymbol{Lemma 1}$\cite{b31}: Define state vector as $\bar y = \left[ {{y_1},{y_2}} \right](\left| {{y_1}} \right| < \min ({L_1}\sin {\theta _1},{L_1}(1 - \sin {\theta _1}),{y_2} \in R)$. Assume that the disturbance is bounded and satisfies $\left| {{{\ddot z}_u}} \right| < \delta  \le {{\zeta {{\left\| {\bar y} \right\|}_2}} \mathord{\left/
		{\vphantom {{\zeta {{\left\| {\bar y} \right\|}_2}} {\sqrt {{{\left( {{{{\mu _1}} \mathord{\left/
											{\vphantom {{{\mu _1}} M}} \right.
											\kern-\nulldelimiterspace} M}} \right)}^2} + 4} }}} \right.
		\kern-\nulldelimiterspace} {\sqrt {{{\left( {{{{\mu _1}} \mathord{\left/
								{\vphantom {{{\mu _1}} M}} \right.
								\kern-\nulldelimiterspace} M}} \right)}^2} + 4} }}$  for all $t \ge 0$, where $\zeta  = \min \left\{ {({{{\mu _1}} \mathord{\left/
							{\vphantom {{{\mu _1}} {{M^2}}}} \right.
						\kern-\nulldelimiterspace} {{M^2}}}){k_v},} \right.\left. {{{(1} \mathord{\left/
					{\vphantom {{(1} {M)}}} \right.
				\kern-\nulldelimiterspace} {M)}}({{{\mu _1}} \mathord{\left/
			{\vphantom {{{\mu _1}} M}} \right.
		\kern-\nulldelimiterspace} M} + 2{\mu _2}{n_x}\vartheta )} \right\}$ and  $\vartheta  = {{({L_1} + {L_2})} \mathord{\left/
	{\vphantom {{({L_1} + {L_2})} {(2{L_1}{L_2}}}} \right.
\kern-\nulldelimiterspace} {(2{L_1}{L_2}}})$, then the bioinspired system \eqref{eq8} will uniformly ultimately bounded.

\section{Approximation-free controller design and analysis}

Before designing the controller, it is necessary to introduce some definitions and theorems, which will be used in subsequent analysis. Given the differential equation and its initial value:

\begin{equation}\label{eq9}
\dot \eta (t) = f(t,\eta (t)),{\rm{ }}\eta (0) = {\eta ^0} \in {\Omega _\eta }.
\end{equation}
where $f:{\Re _ + } \times {\Omega _\xi } \to {\Re ^n}$, and ${\Omega _\xi } \subset {\Re ^n}$ is a non-empty open set. 

$\boldsymbol{Definition 1}$\cite{b34} A solution of the initial value problem \eqref{eq9} (i.e.,$t \mapsto \eta (t)$) is maximal if it has no proper right extension, and it is also a solution for the problem \eqref{eq9}.

$\boldsymbol{Lemma 2}$\cite{b34} Consider the initial value problem \eqref{eq9}. Assume that $f$ is: (a) locally Lipschitz on $\eta $, (b) continuous on $t$ for each fixed $\eta  \in {\Omega _\eta }$ and (c) locally integrable on $t$ for each fixed $\eta  \in {\Omega _\eta }$. Then, a unique maximal solution exits $\eta :\left[ {0,{\tau _{\max }}} \right) \to {\Omega _\eta }$ of \eqref{eq9} on the time interval $\left[ {0,{\tau _{\max }}} \right)$ with ${\tau _{\max }} \in \left\{ {\Re _ + ^ * ,\infty } \right\}$ such that $\eta (t) \in {\Omega _\eta },\forall t \in \left[ {0,{\tau _{\max }}} \right)$.

$\boldsymbol{Proposition 1}$\cite{b34} Assume that the hypotheses of Lemma 2 hold. For a maximal solution $\eta :\left[ {0,{\tau _{\max }}} \right) \to {\Omega _\eta }$ on the time $\left[ {0,{\tau _{\max }}} \right)$ with ${\tau _{\max }} < \infty $ and for any compact set ${\Omega '_\eta } \subset {\Omega _\eta }$, there exists a time instant $t' \in \left[ {0,{\tau _{\max }}} \right)$ such that $\eta (t') \notin {\Omega '_\eta }$.

\subsection{Prescribed performance function}
The tracking error will be transformed to an equivalent error described by a prescribed performance function (PPF). The PPF is a smooth function designed to presuppose the needed performance parameters such as overshoot, the convergence rate, and the ultimate convergence bound of system tracking error. The PPF has the following characteristics: $\rho :{R^ + } \to {R^ + }$, and its value decreases with time. Furthermore, $\mathop {\lim }\limits_{t \to \infty } \rho (t) = {\rho _\infty } > 0$, where ${\rho _\infty }$ is a known constant. Then, the expression $\rho (t) = ({\rho _0} - {\rho _\infty }){e^{ - lt}} + {\rho _\infty }\left( {{\rho _0} > {\rho _\infty } > 0} \right)$ can be one of the possible functions.

Given the upper bound ${\delta _U}$ and lower bound  ${\delta _L}$ and assuming that the tracking error of the system is presented by $e(t)$, then it can be confined to an arbitrarily small set defined by PPF after being controlled, that is
\begin{equation}\label{eq10}
	 - {\delta _L}\rho (t) < e(t) < {\delta _U}\rho (t)
\end{equation}

In order to transform the constraint condition in \eqref{eq10} to unconstraint one, another function $S(\varepsilon )$ is needed, which has the following properties.
(1)	$S(\varepsilon )$ is smooth;
(2)	$- {\delta _L} < S(\varepsilon ) < {\delta _U}$, $\varepsilon  \in ( - \infty , + \infty )$;
(3) $\mathop {\lim }\limits_{\varepsilon  \to  - \infty } S(\varepsilon ) =  - {\delta _L}$,$\mathop {\lim }\limits_{\varepsilon  \to  + \infty } S(\varepsilon ) = {\delta _U}$. The function can be chosen as
\begin{equation}\label{eq11}
S(\varepsilon ) = \frac{{{\delta _U}{e^\varepsilon } - {\delta _L}{e^{ - \varepsilon }}}}{{{e^\varepsilon } + {e^{ - \varepsilon }}}}
\end{equation}
where ${\delta _U}$ and ${\delta _L}$ are positive constant. Then, making $S(\varepsilon ) = {{e(t)} \mathord{\left/
		{\vphantom {{e(t)} {\rho (t)}}} \right.
		\kern-\nulldelimiterspace} {\rho (t)}}$  and combining \eqref{eq10} and \eqref{eq11}, one has that

\begin{equation}\label{eq12}
	\varepsilon  = {S^{ - 1}}(\xi ) = \ln \left( {\frac{{{\delta _L} + \xi }}{{{\delta _U} - \xi }}} \right)
\end{equation}
where $\xi  = {{e(t)} \mathord{\left/
		{\vphantom {{e(t)} {\rho (t)}}} \right.
		\kern-\nulldelimiterspace} {\rho (t)}}$. Hence, the original error $e(t)$ is transformed to equivalent error $\varepsilon$, which varies from minus infinity to infinity. Hence, the control problem can be solved by guaranteeing transformed error $\varepsilon$ bound; then it ensures the error $e(t)$ in \eqref{eq10} can be controlled to a compact set if the parameters $l$, ${\rho _\infty }$, ${\rho _0}$, ${\delta _L}$, and ${\delta _U}$ are chosen properly. 

$\boldsymbol{Remark 1}$ The prescribed performance function  $\rho (t)$ can be designed as asymmetric form based on situations; that is, the prescribed function can consist of upper bound function $\rho {(t)_U}$ and lower bound function $\rho {(t)_L}$, then $\xi $ is described by the following function:
$$\xi  = \frac{{e(t) - \frac{{\rho {{(t)}_U} - \rho {{(t)}_L}}}{2}}}{{\frac{{\rho {{(t)}_U} + \rho {{(t)}_L}}}{2}}}$$

\subsection{Proposed theorem and proof}

Considering the absolute state-space equation \eqref{eq2} and the relative state-space equation \eqref{eq3} of the suspension system that tracks the bioinspired reference model \eqref{eq8}, the approximation-free controller can be designed in three steps. The design procedures are similar to those of backstepping control recursion framework.

$\boldsymbol{Step 1:}$ Define the relative displacement error in \eqref{eq13} and the intermediate virtual control function as follows:

\begin{equation}\label{eq13}
{e_1}(t) = {z_1}(t) - {y_1}(t)
\end{equation}
\begin{equation}\label{eq14}
{u_1} =  - {k_1}{\varepsilon _1} =  - {k_1}\ln \left( {\frac{{{\delta _L}_1 + {\xi _1}}}{{{\delta _{U1}} - {\xi _1}}}} \right)
\end{equation}
where ${k_1}$ is positive control gain; ${\delta _L}_1$, and ${\delta _U}_1$ are positive, and their values are defined by users. ${\xi _1}$ satisfies that ${\xi _1}(t) = {{{e_1}(t)} \mathord{\left/
		{\vphantom {{{e_1}(t)} {{\rho _1}(t)}}} \right.
		\kern-\nulldelimiterspace} {{\rho _1}(t)}}$. ${\rho _1}(t)$ is prescribed performance function, i.e., ${\rho _1}(t) = \left( {{\rho _{10}} - {\rho _{1\infty }}} \right){e^{ - {l_1}t}} + {\rho _{1\infty }}$, where ${\rho _{10}}$ is the initial value and satisfies $\min \left\{ {{\delta _L}_1,{\delta _U}_1} \right\}{\rho _{10}} > \left| {{e_1}(0)} \right| = \left| {{z_1}(0) - {y_1}(0)} \right|$,  ${\rho _{1\infty }}$ is the boundary to make relative displacement error converges to a compact set and ${l_1}$ represents the convergence rate. Users also define their values. 

$\boldsymbol{Step 2:}$ Based on the first intermediate virtual control ${u_1}$  defined in Step 1, the error of relative velocity is given in \eqref{eq15}. The second intermediate virtual control ${u_2}$ can be designed as \eqref{eq16} shown

\begin{equation}\label{eq15}
	{e_2}(t) = {z_2}(t) - {y_2}(t) - {u_1}
\end{equation}

\begin{equation}\label{eq16}
    {u_2} =  - \frac{1}{\theta }{k_2}{\varepsilon _2} =  - \frac{1}{\theta }{k_2}\ln \left( {\frac{{{\delta _{L2}} + {\xi _2}}}{{{\delta _{U2}} - {\xi _2}}}} \right)
\end{equation}
where ${k_2}$, ${\delta _{L2}}$ and ${\delta _{U2}}$ are all positive, and their values are user-defined. Moreover ${\xi _2}(t) = {{{e_2}(t)} \mathord{\left/
		{\vphantom {{{e_2}(t)} {{\rho _2}(t)}}} \right.
		\kern-\nulldelimiterspace} {{\rho _2}(t)}}$ , where ${\rho _2}(t) = \left( {{\rho _{20}} - {\rho _{2\infty }}} \right){e^{ - {l_2}t}} + {\rho _{2\infty }}$. ${\rho _{20}}$, ${\rho _{2\infty }}$ and ${l_2}$ have similar definitions in Step 1. Also, ${\rho _{20}}$ needs to satisfy that $\min \left\{ {{\delta _{L2}},{\delta _{U2}}} \right\}{\rho _{20}} > \left| {{e_2}(0)} \right| = \left| {{z_2}(0) - {y_2}(0) - {u_1}(0)} \right|$.

$\boldsymbol{Step 3:}$ Repeating the same procedure as Step 2, one can define the following absolute velocity error and final controller by virtue of the second virtual control ${u_2}$  
\begin{equation}\label{eq17}
	{e_3}(t) = {x_2}(t) - {u_2}
\end{equation}
\begin{equation}\label{eq18}
	u =  - \frac{1}{\theta }{k_3}{\varepsilon _3} =  - \frac{1}{\theta }{k_3}\ln \left( {\frac{{{\delta _{L3}} + {\xi _3}}}{{{\delta _{U3}} - {\xi _3}}}} \right)
\end{equation}
where ${\xi _3}(t) = {{{e_3}(t)} \mathord{\left/
		{\vphantom {{{e_3}(t)} {{\rho _3}(t)}}} \right.
		\kern-\nulldelimiterspace} {{\rho _3}(t)}}$, ${\rho _3}(t) = \left( {{\rho _{30}} - {\rho _{3\infty }}} \right){e^{ - {l_3}t}} + {\rho _{3\infty }}$ and  $\min \left\{ {{\delta _{L3}},{\delta _{U3}}} \right\}{\rho _{30}} > \left| {{e_3}(0)} \right| = \left| {{x_2}(0) - {u_2}(0)} \right|$. Also, ${\rho _{30}}$, ${\rho _{3\infty }}$, ${l_3}$, ${k_3}$, ${\delta _{L3}}$ and ${\delta _{U3}}$ have similar definition and features as ${\rho _{i0}}$, ${\rho _{i\infty }}$, ${l_i}$, ${k_i}$, ${\delta _{Li}}$ and ${\delta _{Ui}}$ ($i = 1,2$) in Step 1 and Step 2. Additionally, $u$ should satisfy $\left| u \right| > \left| {{u_1} + {u_2}} \right|$. Based on the definition of ${\rho _i}(t),i = 1,2,3$, one can infer that $ - {\delta _L}_i < {\xi _i}(t) = {{{e_i}(t)} \mathord{\left/
		{\vphantom {{{e_i}(t)} {{\rho _i}(t)}}} \right.
		\kern-\nulldelimiterspace} {{\rho _i}(t)}} < {\delta _U}_i,i = 1,2,3$ .
	
$\boldsymbol{Remark 2:}$ Though the approximation-free control based on the bioinspired model has a similar recursion design procedure as backstepping control, the scheme is simpler than backstepping control requiring the adaptive law and the derivative of intermedial virtual control in \cite{b29}, which can be seen from aforementioned. Besides, the simple control scheme avoids the requirement for the knowledge of nonlinearity. Furthermore, the approximators such as FLS or NNs are avoided here compared to \cite{b31,b32,b33}, which reduces the computation complexity.

$\boldsymbol{Remark 3:}$ The suspension system's absolute velocity error is considered in the control design in Step 3. It is known that the acceleration of  the vehicle plays a vital role in riding comfort for human beings. However, the tracking trajectory is dominated by relative states, which means that the design of first and second virtual control only assures convergence of the closed-loop system’s relative signal.  It is necessary to consider the absolute state in the design to confine the absolute acceleration simultaneously, and the final velocity state is a good choice which is the link between absolute displacement state and acceleration one. That is the main aim of Step 3.

Theorem 1 concludes the discussion above.

$\boldsymbol{Theorem 1:}$  Consider the absolute state-space equation \eqref{eq2} and relative one \eqref{eq3} of the suspension system. Assuming the initial state value $\min \left\{ {{\delta _{Li}},{\delta _{Ui}}} \right\}{\rho _{i0}} > \left| {{e_i}(0)} \right|,i = 1,2,3$ fulfilled and given the bioinspired reference model \eqref{eq8}, the transformed error ${\varepsilon _i},i = 1,2,3$ is guaranteed to be bounded under the control scheme \eqref{eq13}-\eqref{eq18} with the condition $\left| u \right| > \left| {{u_1} + {u_2}} \right|$, which retains   within the range $- {\delta _{Li}} < {\xi _i} < {\delta _{Ui}},i = 1,2,3$, so that the tracking error and acceleration of suspension will converge to a small residual set by the assistant of PPF. 

$\boldsymbol{Proof:}$ With the knowledge that ${{{e_i}(t)} \mathord{\left/
		{\vphantom {{{e_i}(t)} {{\rho _i}(t),i = 1,2,3}}} \right.
		\kern-\nulldelimiterspace} {{\rho _i}(t),i = 1,2,3}}$ and considering the expressions \eqref{eq13}-\eqref{eq18}, the state variables in \eqref{eq3} can be presented as ${z_1} = {\xi _1}{\rho _1} + {y_1}$ and ${z_2} = {\xi _2}{\rho _2} + {y_2} + {u_1}$. Then, the derivative of ${\xi _i}(t),i = 1,2,3$ with respect to time can be deduced as follows:

\begin{align}	
	\begin{split}\label{eq19}
	{{\dot \xi }_1} =& {{\left( {{{\dot e}_1}{\rho _1} - {e_1}{{\dot \rho }_1}} \right)} \mathord{\left/
			{\vphantom {{\left( {{{\dot e}_1}{\rho _1} - {e_1}{{\dot \rho }_1}} \right)} {\rho _1^2}}} \right.
			\kern-\nulldelimiterspace} {\rho _1^2}} = {{\left( {\left( {{{\dot z}_1} - {{\dot y}_1}} \right) - {\xi _1}{{\dot \rho }_1}} \right)} \mathord{\left/
			{\vphantom {{\left( {\left( {{{\dot z}_1} - {{\dot y}_1}} \right) - {\xi _1}{{\dot \rho }_1}} \right)} {{\rho _1}}}} \right.
			\kern-\nulldelimiterspace} {{\rho _1}}}\\
	=& {{\left( {{\xi _2}{\rho _2} + {u_1}({\xi _1}) - {\xi _1}{{\dot \rho }_1}} \right)} \mathord{\left/
			{\vphantom {{\left( {{\xi _2}{\rho _2} + {u_1}({\xi _1}) - {\xi _1}{{\dot \rho }_1}} \right)} {{\rho _1}}}} \right.
			\kern-\nulldelimiterspace} {{\rho _1}}} = {\psi _1}(t,{\xi _1},{\xi _2})
    \end{split}\\
    \begin{split}\label{eq20}
  		{{\dot \xi }_2} =& {{\left( {{{\dot e}_2}{\rho _2} - {e_2}{{\dot \rho }_2}} \right)} \mathord{\left/
  				{\vphantom {{\left( {{{\dot e}_2}{\rho _2} - {e_2}{{\dot \rho }_2}} \right)} {\rho _2^2}}} \right.
  				\kern-\nulldelimiterspace} {\rho _2^2}} = {{\left( {\left( {{{\dot z}_2} - {{\dot y}_2} - {{\dot u}_1}} \right) - {\xi _2}{{\dot \rho }_2}} \right)} \mathord{\left/
  				{\vphantom {{\left( {\left( {{{\dot z}_2} - {{\dot y}_2} - {{\dot u}_1}} \right) - {\xi _2}{{\dot \rho }_2}} \right)} {{\rho _2}}}} \right.
  				\kern-\nulldelimiterspace} {{\rho _2}}}\\
  		= &{{\left( {\left( {{{\dot z}_2} - {{\dot y}_2} - {{\dot u}_1}} \right) - {\xi _2}{{\dot \rho }_2}} \right)} \mathord{\left/
  				{\vphantom {{\left( {\left( {{{\dot z}_2} - {{\dot y}_2} - {{\dot u}_1}} \right) - {\xi _2}{{\dot \rho }_2}} \right)} {{\rho _2}}}} \right.
  				\kern-\nulldelimiterspace} {{\rho _2}}}\\
  		=& (\theta \chi \left( {{\xi _1}{\rho _1} + {y_1},{\xi _2}{\rho _2} + {y_2} + {u_1}} \right) + \theta u({\xi _3})\\
  		&{{ + \frac{1}{M}(f({y_1}) + g({y_1},{y_2})) - {{\dot u}_1} - {\xi _2}{{\dot \rho }_2})} \mathord{\left/
  				{\vphantom {{ + \frac{1}{M}(f({y_1}) + g({y_1},{y_2})) - {{\dot u}_1} - {\xi _2}{{\dot \rho }_2})} {{\rho _2}}}} \right.
  				\kern-\nulldelimiterspace} {{\rho _2}}}\\
  		=& {\psi _2}(t,{\xi _1},{\xi _2},{\xi _3})
    \end{split}\\
    \begin{split}\label{eq21}
	{{\dot \xi }_3} =& {{\left( {{{\dot e}_3}{\rho _3} - {e_3}{{\dot \rho }_3}} \right)} \mathord{\left/
			{\vphantom {{\left( {{{\dot e}_3}{\rho _3} - {e_3}{{\dot \rho }_3}} \right)} {\rho _3^2}}} \right.
			\kern-\nulldelimiterspace} {\rho _3^2}} = {{\left( {\left( {{{\dot x}_2} - {{\dot u}_2}} \right) - {\xi _3}{{\dot \rho }_3}} \right)} \mathord{\left/
			{\vphantom {{\left( {\left( {{{\dot x}_2} - {{\dot u}_2}} \right) - {\xi _3}{{\dot \rho }_3}} \right)} {{\rho _3}}}} \right.
			\kern-\nulldelimiterspace} {{\rho _3}}}\\
	=& \left( {\theta \chi \left( {{\xi _1}{\rho _1} + {y_1},{\xi _2}{\rho _2} + {y_2} + {u_1}} \right) + \theta u({\xi _3})} \right.\\
	&{{\left. { - {{\dot u}_2} - {\xi _3}{{\dot \rho }_3}} \right)} \mathord{\left/
			{\vphantom {{\left. { - {{\dot u}_2} - {\xi _3}{{\dot \rho }_3}} \right)} {{\rho _3}}}} \right.
			\kern-\nulldelimiterspace} {{\rho _3}}}\\
	=& {\psi _3}(t,{\xi _1},{\xi _2},{\xi _3})
    \end{split}
\end{align}
\eqref{eq19}-\eqref{eq21} can be further expressed by
\begin{equation}\label{eq22}
	\dot \xi  = \psi (t,\xi ) = \left[ \begin{array}{l}
		{\psi _1}(t,{\xi _1},{\xi _2})\\
		{\psi _2}(t,{\xi _1},{\xi _2},{\xi _3})\\
		{\psi _3}(t,{\xi _1},{\xi _2},{\xi _3})
	\end{array} \right]
\end{equation}

There are two phases to complete the proof. The first phase ensures the existence and uniqueness of a maximal solution $\xi :[0,{\tau _{\max }}) \to {\Omega _\xi }$ of \eqref{eq22}, where ${\Omega _\xi } = \left( { - {\delta _{L1}},{\delta _{U1}}} \right) \times \left( { - {\delta _{L2}},{\delta _{U2}}} \right) \times \left( { - {\delta _{L3}},{\delta _{U3}}} \right)$. Secondly, the transformed errors  ${\varepsilon _i},i = 1,2,3$ need to be bounded under the proposed controller over the time interval $t \in \left( {0,{\tau _{\max }}} \right)$, which ensures the system stability. Subsequently, the theorem is extended to the global one by proving ${\tau _{\max }} =  + \infty $.

$\boldsymbol{Phase A}$ With the condition $\min \left\{ {{\delta _{Li}},{\delta _{Ui}}} \right\}{\rho _{i0}} > \left| {{e_i}(0)} \right|,i = 1,2,3$, it’s clear that the initial value of $\xi $ satisfies $\left| {{\xi _i}(0)} \right| < \min \left\{ {{\delta _{Li}},{\delta _{Ui}}} \right\},i = 1,2,3$ , i.e., $\xi (0) \in {\Omega _\xi }$. Besides, the state vector $[{y_1},{y_2}]$ of bioinspired reference model has already been proved bounded in Section 2. Furthermore, the suspension system and prescribed performance function ${\rho _i},i = 1,2,3$ are continuous and differentiable. Then $\psi (t,\xi )$ in \eqref{eq22} is continuous and differentiable and satisfies the locally Lipschitz condition, which implies Lemma 2 can be applied here. Hence, there exists a maximal solution $\xi :[0,{\tau _{\max }}) \to {\Omega _\xi }$ such that $\xi (t) \in {\Omega _\xi },\forall t \in [0,{\tau _{\max }})$, that is, $\left| {{\xi _i}(t)} \right| < \min \left\{ {{\delta _{Li}},{\delta _{Ui}}} \right\},i = 1,2,3,\forall t \in [0,{\tau _{\max }})$.

$\boldsymbol{Phase B}$ Based on the definition of the prescribed performance function and the characteristic of the designed controller, we know that system's stability is guaranteed if the transformed error ${\varepsilon _i},i = 1,2,3$ remains bounded. For transformed error  ${\varepsilon _1} = \ln \left( {{{\left( {{\delta _L}_1 + {\xi _1}} \right)} \mathord{\left/
			{\vphantom {{\left( {{\delta _L}_1 + {\xi _1}} \right)} {\left( {{\delta _{U1}} - {\xi _1}} \right)}}} \right.
			\kern-\nulldelimiterspace} {\left( {{\delta _{U1}} - {\xi _1}} \right)}}} \right)$, the following Lyapunov function \eqref{eq23} is constructed. 		
\begin{equation}\label{eq23}
	{V_1} = \frac{1}{2}\varepsilon _1^2
\end{equation}
Taking the derivative of equation \eqref{eq23} for $t \in [0,{\tau _{\max }})$, and substituting equation \eqref{eq19}, one has that
\begin{equation}\label{eq24}
	\begin{split}
		{{\dot V}_1} =& {\varepsilon _1}{{\dot \varepsilon }_1}\\
		=& {\varepsilon _1}\left( {\frac{{{\delta _{U1}} + {\delta _{L1}}}}{{\left( {{\delta _{L1}} + {\xi _1}} \right)\left( {{\delta _{U1}} - {\xi _1}} \right)}}{{\dot \xi }_1}} \right)\\
		=& \frac{{{\delta _{U1}} + {\delta _{L1}}}}{{\left( {{\delta _{L1}} + {\xi _1}} \right)\left( {{\delta _{U1}} - {\xi _1}} \right){\rho _1}}}{\varepsilon _1}\left( {{{\dot z}_1} - {{\dot y}_1} - {\xi _1}{{\dot \rho }_1}} \right)
	\end{split}
\end{equation}

Making ${r_1} = {{\left( {{\delta _{U1}} + {\delta _{L1}}} \right)} \mathord{\left/
		{\vphantom {{\left( {{\delta _{U1}} + {\delta _{L1}}} \right)} {\left( {\left( {{\delta _{L1}} + {\xi _1}} \right)\left( {{\delta _{U1}} - {\xi _1}} \right){\rho _1}} \right)}}} \right.
		\kern-\nulldelimiterspace} {\left( {\left( {{\delta _{L1}} + {\xi _1}} \right)\left( {{\delta _{U1}} - {\xi _1}} \right){\rho _1}} \right)}}$. With the knowledge that ${\delta _{U1}}$, and ${\delta _{L1}}$ are positive and bounded, ${\rho _1}$, and ${\rho _2}$ are positive continuous bounded functions, and ${\xi _1}$ satisfies $ - {\delta _{L1}} < {\xi _1} < {\delta _{U1}}$, then it can be inferred that ${r_1} > 0$. Taking account into ${\dot z_1} - {\dot y_1} = {z_2} - {y_2} = {e_2} + {u_1}$, it follows from \eqref{eq24} that
\begin{equation}\label{eq25}
\begin{split}
	{{\dot V}_1} &= {r_1}{\varepsilon _1}\left( {{e_2} + {u_1} - {\xi _1}{{\dot \rho }_1}} \right) = {r_1}{\varepsilon _1}\left( {{\xi _2}{\rho _2} - {k_1}{\varepsilon _1} - {\xi _1}{{\dot \rho }_1}} \right)\\
	&\le {r_1}\left( {\left| {{\varepsilon _1}} \right|{{\bar Q}_1} - {k_1}{{\left| {{\varepsilon _1}} \right|}^2}} \right)
\end{split}	
\end{equation}
where ${\bar Q_1}$ is a positive constant representing bound of $\left| {{\xi _2}{\rho _2} - {\xi _1}{{\dot \rho }_1}} \right|$. We know that ${\rho _1}$ is continuous and monotonically decreasing, thus ${\dot \rho _1}$ is bounded. Besides, ${\xi _2}$ is bounded and satisfies that $ - {\delta _{L2}} < {\xi _2} < {\delta _{U2}}$ . Therefore, the conclusion is that $\left| {{\xi _2}{\rho _2} - {\xi _1}{{\dot \rho }_1}} \right|$ is bounded based on the Extreme Value Theorem, then it is represented by the supremum ${\bar Q_1}$, i.e., ${\bar Q_1} = \sup \left\{ {\left| {{\xi _2}{\rho _2} - {\xi _1}{{\dot \rho }_1}} \right|} \right\}$ .
Hence, ${\dot V_1}$ is negative if ${\varepsilon _1}$ satisfies the condition $\left| {{\varepsilon _1}} \right| > {{{{\bar Q}_1}} \mathord{\left/
		{\vphantom {{{{\bar Q}_1}} {{k_1}}}} \right.
		\kern-\nulldelimiterspace} {{k_1}}}$. Consequently, we have the conclusion that the transformed error ${\varepsilon _1}$ will converge to compact set ${\Omega _1} = \left\{ {{\varepsilon _1}\left| {\left| {{\varepsilon _1}} \right| < {{\bar \varepsilon }_1}} \right.} \right\}$ with ${\bar \varepsilon _1} = \max \left\{ {\left| {{\varepsilon _1}(0)} \right|,{{{{\bar Q}_1}} \mathord{\left/
			{\vphantom {{{{\bar Q}_1}} {{k_1}}}} \right.
			\kern-\nulldelimiterspace} {{k_1}}}} \right\},\forall t \in [0,{\tau _{\max }})$, based on Lyapunov Theorem. In addition, the first intermediate virtual control $ {u_1} =  - {k_1}{\varepsilon _1}$ is bounded.

Next, the similar Lyapunov function is given to prove the second transformed error ${\varepsilon _2} = \ln \left( {{{\left( {{\delta _L}_2 + {\xi _2}} \right)} \mathord{\left/
			{\vphantom {{\left( {{\delta _L}_2 + {\xi _2}} \right)} {\left( {{\delta _{U2}} - {\xi _2}} \right)}}} \right.
			\kern-\nulldelimiterspace} {\left( {{\delta _{U2}} - {\xi _2}} \right)}}} \right)$ as follows:
\begin{equation}\label{eq26}
{V_2} = \frac{1}{2}\varepsilon _2^2	
\end{equation}	
Then for $t \in [0,{\tau _{\max }})$ the following derivative of \eqref{eq26} can be obtained by substituting  \eqref{eq20}, \eqref{eq3} and \eqref{eq8}
\begin{equation}\label{eq27}
	\begin{split}
	{{\dot V}_2} =& {\varepsilon _2}{{\dot \varepsilon }_2}\\
	=& {\varepsilon _2}\left( {\frac{{{\delta _{U2}} + {\delta _{L2}}}}{{\left( {{\delta _{L2}} + {\xi _2}} \right)\left( {{\delta _{U2}} - {\xi _2}} \right)}}{{\dot \xi }_2}} \right)\\
	=& \frac{{{\delta _{U2}} + {\delta _{L2}}}}{{\left( {{\delta _{L2}} + {\xi _2}} \right)\left( {{\delta _{U2}} - {\xi _2}} \right){\rho _2}}}{\varepsilon _2}\left( {{{\dot z}_2} - {{\dot y}_2} - {{\dot u}_1} - {\xi _2}{{\dot \rho }_2}} \right)\\
	=& {r_2}{\varepsilon _2}\left( {\theta \chi \left( {{z_1},{z_2}} \right) + \theta u + \left( {{1 \mathord{\left/
					{\vphantom {1 M}} \right.
					\kern-\nulldelimiterspace} M}} \right)(f({y_1}) + g({y_1},{y_2})} \right.\\
	&\left. { - {{\dot u}_1} - {\xi _2}{{\dot \rho }_2}} \right)
	\end{split}	
\end{equation}
where ${r_2} = {{\left( {{\delta _{U2}} + {\delta _{L2}}} \right)} \mathord{\left/
		{\vphantom {{\left( {{\delta _{U2}} + {\delta _{L2}}} \right)} {\left( {\left( {{\delta _{L2}} + {\xi _2}} \right)\left( {{\delta _{U2}} - {\xi _2}} \right){\rho _2}} \right)}}} \right.
		\kern-\nulldelimiterspace} {\left( {\left( {{\delta _{L2}} + {\xi _2}} \right)\left( {{\delta _{U2}} - {\xi _2}} \right){\rho _2}} \right)}}$. With the condition that ${\delta _{U2}}$, and ${\delta _{L2}}$ are positive and  satisfies that $ - {\delta _{L2}} < {\xi _2} < {\delta _{U2}}$, then $\left( {{\delta _{L2}} + {\xi _2}} \right)\left( {{\delta _{U2}} - {\xi _2}} \right) > 0$. Besides, ${\rho _2}$ is positive. Therefore, it is deduced that ${r_2}$ is positive. Assuming that $\left| u \right| > \left| {{u_1} + {u_2}} \right|$ and introducing the condition ${z_1} = {\xi _1}{\rho _1} + {y_1}$ and ${z_2} = {\xi _2}{\rho _2} + {y_2} + {u_1}$, it can be further derived from \eqref{eq27} that 
\begin{equation}\label{eq28}
	\begin{split}
	{{\dot V}_2} \le& {r_2}{\varepsilon _2}(\theta \chi \left( {{\xi _1}{\rho _1} + {y_1},{\xi _2}{\rho _2} + {y_2} + {u_1}} \right) + \theta \left( {{u_1} + {u_2}} \right)\\
	&+ \frac{1}{M}(f({y_1}) + g({y_1},{y_2})) - {{\dot u}_1} - {\xi _2}{{\dot \rho }_2})\\
	=& {r_2}{\varepsilon _2}(\theta \chi \left( {{\xi _1}{\rho _1} + {y_1},{\xi _2}{\rho _2} + {y_2} + {u_1}} \right) + \theta {u_1} - {k_2}{\varepsilon _2}\\
	&+ \frac{1}{M}(f({y_1}) + g({y_1},{y_2})) - {{\dot u}_1} - {\xi _2}{{\dot \rho }_2})\\
	\le& {r_2}\left( {\left| {{\varepsilon _2}} \right|{{\bar Q}_2} - {k_2}{{\left| {{\varepsilon _2}} \right|}^2}} \right)
	\end{split}	
\end{equation}
where ${\bar Q_2} = \sup \left\{ {\left| {\theta \chi  + \theta {u_1} + {{(1} \mathord{\left/
				{\vphantom {{(1} {{{M)} \mathord{\left/
									{\vphantom {{M)} {(f + g) - {{\dot u}_1} - {\xi _2}{{\dot \rho }_2}}}} \right.
									\kern-\nulldelimiterspace} {(f + g) - {{\dot u}_1} - {\xi _2}{{\dot \rho }_2}}}}}} \right.
				\kern-\nulldelimiterspace} {{{M)} \mathord{\left/
						{\vphantom {{M)} {(f + g) - {{\dot u}_1} - {\xi _2}{{\dot \rho }_2}}}} \right.
						\kern-\nulldelimiterspace} {(f + g) - {{\dot u}_1} - {\xi _2}{{\dot \rho }_2}}}}}} \right|} \right\}$. For simplicity, $\chi \left( {{\xi _1}{\rho _1} + {y_1},{\xi _2}{\rho _2} + {y_2} + {u_1}} \right)$, $f({y_1})$, and $g({y_1},{y_2})$ in \eqref{eq28} are abbreviated by $\chi$, $f$, and $g$, respectively. According to lemma 1, the bioinspired model \eqref{eq8} is uniformly ultimately bounded, which means ${y_1}$ ,${y_2}$ and ${{(1} \mathord{\left/
				{\vphantom {{(1} {{{M)} \mathord{\left/
						{\vphantom {{M)} {(f + g)}}} \right.
						\kern-\nulldelimiterspace} {(f + g)}}}}} \right.
			\kern-\nulldelimiterspace} {{{M)} \mathord{\left/
		{\vphantom {{M)} {(f + g)}}} \right.
	\kern-\nulldelimiterspace} {(f + g)}}}}$ are bounded. With the assumption that ${F_s}$ and ${F_d}$ are continuous and bound, we know that $\chi  =  - {F_d} - {F_s}$ is bounded. Besides, ${u_1}$ and ${\rho _2}$ are bounded and continuous, which indicates that ${\dot u_1}$, and ${\dot \rho _2}$ are also bounded. Consequently, applying the Extreme Value Theorem, $\left| {\theta \chi  + \theta {u_1} + {{(1} \mathord{\left/
{\vphantom {{(1} {{{M)} \mathord{\left/
				{\vphantom {{M)} {(f + g) - {{\dot u}_1} - {\xi _2}{{\dot \rho }_2}}}} \right.
				\kern-\nulldelimiterspace} {(f + g) - {{\dot u}_1} - {\xi _2}{{\dot \rho }_2}}}}}} \right.
\kern-\nulldelimiterspace} {{{M)} \mathord{\left/
	{\vphantom {{M)} {(f + g) - {{\dot u}_1} - {\xi _2}{{\dot \rho }_2}}}} \right.
	\kern-\nulldelimiterspace} {(f + g) - {{\dot u}_1} - {\xi _2}{{\dot \rho }_2}}}}}} \right|$ has an upper bound, denoted by ${\bar Q_2}$.
As a consequence, ${\dot V_2}$ is negative if $\left| {{\varepsilon _2}} \right| > {{{{\bar Q}_2}} \mathord{\left/
		{\vphantom {{{{\bar Q}_2}} {{k_2}}}} \right.
		\kern-\nulldelimiterspace} {{k_2}}}$ and according to Lyapunov Theorem, ${\varepsilon _2}$ will converge to compact set ${\Omega _2} = \left\{ {{\varepsilon _2}\left| {\left| {{\varepsilon _2}} \right| < {{\bar \varepsilon }_2}} \right.} \right\}$ with $ {\bar \varepsilon _2} = \max \left\{ {\left| {{\varepsilon _2}(0)} \right|,{{{{\bar Q}_2}} \mathord{\left/
			{\vphantom {{{{\bar Q}_2}} {{k_2}}}} \right.
			\kern-\nulldelimiterspace} {{k_2}}}} \right\},\forall t \in [0,{\tau _{\max }})$. Then, the second intermediate virtual control ${u_2} =  - {k_2}{\varepsilon _2}$ is also guaranteed to be bounded.

$\boldsymbol{Remark 4}$ The reason for using the condition $\left| u \right| > \left| {{u_1} + {u_2}} \right|$ is presented here. It is easy to know that ${\varepsilon _i},i = 1,2,3$ have opposite sigh against ${u_i},i = 1,2,3{\rm{ }}(u = {u_3})$ according to their expressions. As mentioned above, one has that ${\varepsilon _3} = \ln \left( {{{\left( {{\delta _{L3}} + {\xi _3}} \right)} \mathord{\left/
			{\vphantom {{\left( {{\delta _{L3}} + {\xi _3}} \right)} {\left( {{\delta _{U3}} - {\xi _3}} \right)}}} \right.
			\kern-\nulldelimiterspace} {\left( {{\delta _{U3}} - {\xi _3}} \right)}}} \right)$ and ${\xi _3} = {{\left( {{x_2} + {k_2}{\varepsilon _2}} \right)} \mathord{\left/
		{\vphantom {{\left( {{x_2} + {k_2}{\varepsilon _2}} \right)} {{\rho _3}}}} \right.
	\kern-\nulldelimiterspace} {{\rho _3}}}$. The partial differential of ${\varepsilon _3}$ with respect to ${\xi _3}$ and the partial differential of ${\xi _3}$ with respect to ${\varepsilon _2}$ is positive according to ${k_2} > 0,{\rho _3} > 0$, and ${{\left( {{\delta _{U2}} + {\delta _{L2}}} \right)} \mathord{\left/
	{\vphantom {{\left( {{\delta _{U2}} + {\delta _{L2}}} \right)} {\left( {\left( {{\delta _{L2}} + {\xi _2}} \right)\left( {{\delta _{U2}} - {\xi _2}} \right)} \right)}}} \right.
	\kern-\nulldelimiterspace} {\left( {\left( {{\delta _{L2}} + {\xi _2}} \right)\left( {{\delta _{U2}} - {\xi _2}} \right)} \right)}} > 0$. Besides, if ${\varepsilon _2} \to 0$, then ${\xi _3} \to 0$ is satisfied due to the expressions of ${\varepsilon _2}$ and ${\xi _3}$. Also, if ${\xi _3} \to 0$ then ${\varepsilon _3} \to 0$ is required based on the control goal. Therefore, ${\varepsilon _2}$ has the same sign as ${\varepsilon _3}$ and $u$ has a sign opposite of ${\varepsilon _2}$. By analogy, one can conclude that ${\varepsilon _i},i = 1,2,3$ and ${u_i},i = 1,2,3(u = {u_3})$ have the same sign with each other, respectively, but ${\varepsilon _i},i = 1,2,3$ have opposite sign against ${u_i},i = 1,2,3(u = {u_3})$, which helps the use of the condition $\left| u \right| > \left| {{u_1} + {u_2}} \right|$ in \eqref{eq28}.

In the same manner, the similar Lyapunov structure is presented for the third transformed error ${\varepsilon _3} = \ln \left( {{{\left( {{\delta _L}_3 + {\xi _3}} \right)} \mathord{\left/
			{\vphantom {{\left( {{\delta _L}_3 + {\xi _3}} \right)} {\left( {{\delta _{U3}} - {\xi _3}} \right)}}} \right.
			\kern-\nulldelimiterspace} {\left( {{\delta _{U3}} - {\xi _3}} \right)}}} \right)$
\begin{equation}\label{eq29}
{V_3} = \frac{1}{2}\varepsilon _3^2
\end{equation}
Introducing \eqref{eq21}, the derivative of \eqref{eq29} can be deduced as follows:	
\begin{equation}\label{eq30}
\begin{split}
	{{\dot V}_3} =& {\varepsilon _3}{{\dot \varepsilon }_3}\\
	=& {\varepsilon _3}\left( {\frac{{{\delta _{U3}} + {\delta _{L3}}}}{{\left( {{\delta _{L3}} + {\xi _3}} \right)\left( {{\delta _{U3}} - {\xi _3}} \right)}}{{\dot \xi }_3}} \right)\\
	=& \frac{{{\delta _{U3}} + {\delta _{L3}}}}{{\left( {{\delta _{L3}} + {\xi _3}} \right)\left( {{\delta _{U3}} - {\xi _3}} \right){\rho _3}}}{\varepsilon _3}\left( {{{\dot x}_2} - {{\dot u}_2} - {\xi _3}{{\dot \rho }_3}} \right)
\end{split}
\end{equation}	
If ${r_3} = {{\left( {{\delta _{U3}} + {\delta _{L3}}} \right)} \mathord{\left/
		{\vphantom {{\left( {{\delta _{U3}} + {\delta _{L3}}} \right)} {\left( {\left( {{\delta _{L3}} + {\xi _3}} \right)\left( {{\delta _{U3}} - {\xi _3}} \right){\rho _3}} \right)}}} \right.
		\kern-\nulldelimiterspace} {\left( {\left( {{\delta _{L3}} + {\xi _3}} \right)\left( {{\delta _{U3}} - {\xi _3}} \right){\rho _3}} \right)}}$ then ${r_3} > 0$ as $ - {\delta _{L3}} < {\xi _3} < {\delta _{U3}}$, where ${\delta _{U3}}$, ${\delta _{L3}}$ and ${\rho _3}$ are positive. Substituting absolute state-space form \eqref{eq2} of the suspension system, the expression \eqref{eq30} can be further derived into the following equation 
\begin{equation}\label{eq31}
\begin{split}
	{{\dot V}_3} =& {r_3}{\varepsilon _3}\left( {\theta (\chi ({z_1},{z_2}) + u) - {{\dot u}_2} - {\xi _3}{{\dot \rho }_3}} \right)\\
	=& {r_3}{\varepsilon _3}(\theta \chi ({\xi _2}{\rho _2} + {y_2} + {u_1},{\xi _1}{\rho _1} + {y_1}) - {k_3}{\varepsilon _3} \\
	&- {{\dot u}_2} - {\xi _3}{{\dot \rho }_3})\\
	\le& {r_3}\left( {\left| {{\varepsilon _3}} \right|{{\bar Q}_3} - {k_3}{{\left| {{\varepsilon _3}} \right|}^2}} \right)
\end{split}
\end{equation}
where ${\bar Q_3} = \left\{ {\left| {\theta \chi ({\xi _2}{\rho _2} + {y_2} + {u_1},{\xi _1}{\rho _1} + {y_1}) - {{\dot u}_2} - {\xi _3}{{\dot \rho }_3}} \right|} \right\}$. Because ${u_2}$ and ${\rho _3}$ are bounded and continuous, then ${\dot u_2}$ and ${\dot \rho _3}$ are all bounded. Consequently, according to the Extreme Value Theorem, there exists a supremum for $\left| {\theta \chi ({\xi _2}{\rho _2} + {y_2} + {u_1},{\xi _1}{\rho _1} + {y_1}) - {{\dot u}_2} - {\xi _3}{{\dot \rho }_3}} \right|$, which is represented by ${\bar Q_3}$

Therefore, ${\dot V_3}$ is negative for any $\left| {{\varepsilon _3}} \right| > {{{{\bar Q}_3}} \mathord{\left/
		{\vphantom {{{{\bar Q}_3}} {{k_3}}}} \right.
		\kern-\nulldelimiterspace} {{k_3}}}$. Then it is deduced from Lyapunov Theorem  that ${\varepsilon _3}$ will converged to a compact set ${\Omega _3} = \left\{ {{\varepsilon _3}\left| {\left| {{\varepsilon _3}} \right| < {{\bar \varepsilon }_3}} \right.} \right\}$ with ${\bar \varepsilon _3} = \max \left\{ {\left| {{\varepsilon _3}(0)} \right|,{{{{\bar Q}_3}} \mathord{\left/
			{\vphantom {{{{\bar Q}_3}} {{k_3}}}} \right.
			\kern-\nulldelimiterspace} {{k_3}}}} \right\},\forall t \in [0,{\tau _{\max }})$. Finally, it is concluded that the control force $u$ is bounded. Consequently, the relative states ${z_1}$, ${z_2}$ and absolute state ${x_2}$ remain bounded. Moreover, one can obtain that
\begin{equation}\label{eq32}
	\begin{split}
	&- {\delta _{Li}} < \frac{{{e^{ - {{\bar \varepsilon }_i}}}{\delta _{Ui}} - {\delta _{Li}}}}{{{e^{ - {{\bar \varepsilon }_i}}} + 1}} = {\xi _{Li}} \le {\xi _i}\\
	&{\xi _i} \le {\xi _{Ui}} = \frac{{{e^{{{\bar \varepsilon }_i}}}{\delta _{Ui}} - {\delta _{Li}}}}{{{e^{{{\bar \varepsilon }_i}}} + 1}} < {\delta _{Ui}},i = 1,2,3,\forall t \in [0,{\tau _{\max }})
   \end{split}
\end{equation}

Defining the set ${\Omega '_\xi } = [{\xi _{L1}},{\xi _{U1}}] \times [{\xi _{L2}},{\xi _{U2}}] \times [{\xi _{L3}},{\xi _{U3}}]$, it is obvious that ${\Omega '_\xi } < {\Omega _\xi }$ from \eqref{eq32}.  Assuming that ${\tau _{\max }} <  + \infty $, we know that there exists a time instant $t' \in [0,{\tau _{\max }})$ such that ${\xi _i}(t') \notin {\Omega '_\xi }$ by applying Proposition 1, which is contradictory to what is expressed in \eqref{eq32}. Thus, we have the conclusion that ${\tau _{\max }} =  + \infty $ and Theorem 1 is applicable for $t \in [0, + \infty )$. The proof of Theorem 1 is complete.

\subsection{Suspension ride safety analysis}
Additionally, the suspension system needs to meet the requirements of ride safety mentioned in Section 2 and it will be analyzed next. From the above proof of Theorem 1, we know that relative displacement ${z_1} = {x_1} - {x_3}$ of sprung mass against unsprung mass can be confined to a reasonable residual set, that is, $ - {\delta _{L1}}{\rho _{10}} < {z_1} - {y_1} < {\delta _{U1}}{\rho _{10}}$. It implies that the suspension deflection constraint \eqref{eq5} is guaranteed if parameters of controller ${\delta _{L1}}$, ${\delta _{U1}}$, and ${\rho _{10}}$ are chosen appropriately, as well as parameters of bioinspired reference model. The bioinspired reference model \eqref{eq3} is a kind of vibration isolator and provide an ideal reference trajectory to controller. One can refer to Ref. \cite{b30} to choose appropriate parameters of bioinspired model. For dynamic tire load, the analysis is presented next. Considering state $[{x_3},{x_4}]$, the following equation holds
\begin{equation}\label{eq33}
x = Ax + w
\end{equation}
where
$$A = \left[ {\begin{array}{*{20}{c}}
		0&1\\
		{ - \frac{{{k_t}}}{{{m_u}}}}&{ - \frac{{{c_t}}}{{{m_u}}}}
\end{array}} \right]$$ and $$w = \left[ {\begin{array}{*{20}{c}}
0\\
{\frac{1}{{{m_u}}}( - {k_t}{z_r} - {c_t}{{\dot z}_r} + {F_d} + {F_s} - u)}
\end{array}} \right]$$

Based on previous proof of theorem 1, it's known that $u$ is bounded. Combining Assumption 2 and 3, it can be inferred that $w$ has the limit, which is presented by $w \le \bar w$. For the system \eqref{eq33}, the Lyapunov function ${V_x} = {x^T}Px$ is constructed, where $P > 0$ is a symmetric matrix. Thus, its derivative can be obtained as ${\dot V_x} = {x^T}\left( {{A^T}P + PA} \right)x + 2{w^T}Px$. Moreover, the condition ${A^T}P + PA =  - Q$ can be utilized since $A$ is Hurwitz. Then Young’s inequality can be applied in \eqref{eq33}. Therefore, the derivative is further deduced by
\begin{equation}\label{eq34}
	\begin{split}
	{{\dot V}_x} &\le  - {x^T}Qx + \frac{{{\lambda _{\max }}({P^2})}}{\eta }{\left\| x \right\|^2} + \eta {\left\| w \right\|^2}\\
	&\le  - \left( {{\lambda _{\min }}(Q) - {{{\lambda _{\max }}({P^2})} \mathord{\left/
				{\vphantom {{{\lambda _{\max }}({P^2})} \eta }} \right.
				\kern-\nulldelimiterspace} \eta }} \right){\left\| x \right\|^2} + \eta {{\bar w}^2}\\
	&\le  - \ell {V_x} + W
    \end{split}
\end{equation}
where $\ell  = {\lambda _{\min }}(Q) - {{{\lambda _{\max }}({P^2})} \mathord{\left/
		{\vphantom {{{\lambda _{\max }}({P^2})} \eta }} \right.
		\kern-\nulldelimiterspace} \eta }$, $W = \eta {\bar w^2}$ and $\eta $ is a positive constant defined in Young’s inequality. Integrating both  sides of \eqref{eq34}, one can obtain that
\begin{equation}\label{eq35}
	{V_x}(t) \le {V_x}(0){e^{ - \ell t}} + {W \mathord{\left/
			{\vphantom {W {\ell  \le }}} \right.
			\kern-\nulldelimiterspace} {\ell  \le }}{V_x}(0) + {W \mathord{\left/
			{\vphantom {W \ell }} \right.
			\kern-\nulldelimiterspace} \ell }
\end{equation}

Therefore, from \eqref{eq35} it implies that suspension states ${x_3}$ and ${x_4}$ are bounded, that is, $\left| {{x_i}} \right| \le \sqrt {{{\left( {{V_x}(0) + {W \mathord{\left/
						{\vphantom {W \ell }} \right.
						\kern-\nulldelimiterspace} \ell }} \right)} \mathord{\left/
			{\vphantom {{\left( {{V_x}(0) + {W \mathord{\left/
									{\vphantom {W \ell }} \right.
									\kern-\nulldelimiterspace} \ell }} \right)} {{\lambda _{\min }}\left( P \right)}}} \right.
			\kern-\nulldelimiterspace} {{\lambda _{\min }}\left( P \right)}}} ,{\rm{  }}i = 3,4$. Furthermore, based on Assumption 3, the following inequality holds
\begin{equation}\label{eq36}
	\begin{split}
	\left| {{F_t} + {F_b}} \right| =& \left| {{k_t}\left( {{x_3} - {z_r}} \right) + {c_t}\left( {{x_3} - {z_r}} \right)} \right|\\
	\le& \left( {{k_t} + {c_t}} \right)\sqrt {{{\left( {{V_x}(0) + {W \mathord{\left/
							{\vphantom {W \ell }} \right.
							\kern-\nulldelimiterspace} \ell }} \right)} \mathord{\left/
				{\vphantom {{\left( {{V_x}(0) + {W \mathord{\left/
										{\vphantom {W \ell }} \right.
										\kern-\nulldelimiterspace} \ell }} \right)} {{\lambda _{\min }}\left( P \right)}}} \right.
				\kern-\nulldelimiterspace} {{\lambda _{\min }}\left( P \right)}}} \\
	&+ {k_t}{{\bar z}_{r1}} + {c_t}{{\bar z}_{r2}}
	\end{split}
\end{equation}
Consequently, the dynamic tyre load constraint \eqref{eq4} is guaranteed by designing appropriate parameters $\eta $ and $P$ that satisfy the inequality $\left( {{k_t} + {c_t}} \right) \cdot \sqrt {{{\left( {{V_x}(0) + {W \mathord{\left/
						{\vphantom {W \ell }} \right.
						\kern-\nulldelimiterspace} \ell }} \right)} \mathord{\left/
			{\vphantom {{\left( {{V_x}(0) + {W \mathord{\left/
									{\vphantom {W \ell }} \right.
									\kern-\nulldelimiterspace} \ell }} \right)} {{\lambda _{\min }}\left( P \right)}}} \right.
			\kern-\nulldelimiterspace} {{\lambda _{\min }}\left( P \right)}}}  + {k_t}{\bar z_{r1}} + {c_t}{\bar z_{r2}} \le \left( {{m_u} + {m_s}} \right)g$, which ensures ride holding performance of the suspension system.

\subsection{Theoretical analysis in PPF superior convergence property}		
The property of preset transient performance of approximation-free controller makes it superior to the controller with FLS in convergence rate. In this part, the better convergence performance of the approximation-free controller will be analyzed theoretically.

Firstly, we will prove that the system controlled by FLS-controller with PPF converges faster than the one controlled by FLS-controller without PPF[29, 31], which confirms the effectiveness of transient property predefined by PPF. Before presenting the proof, the following inequality will be needed:
\begin{equation}\label{eq37}
\left| {\ln \left( {\frac{{a + \xi }}{{b - \xi }}} \right)} \right| \ge \frac{4}{{a + b}}\left| {\xi  - \frac{{b - a}}{2}} \right|
\end{equation}
where $a > 0$, $b > 0$, and $ - a < \xi  < b$. The inequality \eqref{eq37} can be easily proved by dividing it into two parts: one fo $\xi  \in \left\{ {\left. \xi  \right|\xi  > {{\left( {b - a} \right)} \mathord{\left/
			{\vphantom {{\left( {b - a} \right)} 2}} \right.
			\kern-\nulldelimiterspace} 2}} \right\}$, another for $\xi  \in \left\{ {\left. \xi  \right|\xi  \le } \right.\left. {{{\left( {b - a} \right)} \mathord{\left/
			{\vphantom {{\left( {b - a} \right)} 2}} \right.
			\kern-\nulldelimiterspace} 2}} \right\}$, and taking the derivative.

Here the suspension system \eqref{eq3} tracking bioinspired model \eqref{eq8} is considered.  Define the system state as $[{x_{s1}},{x_{s2}}] = [{z_1} - {y_1},{z_2} - {y_2}]$, error as $[{e_1},{e_2}] = [{x_{s1}},{x_{s2}} - {\alpha _1}]$,  and intermediate virtual control as ${\alpha _1} =  - {\lambda _1}{e_1}$, then the fuzzy control law and adaptive law can be designed as:
\begin{align}
\label{eq38}
&u = \frac{1}{\theta }( - {\lambda _2}{e_2} - {e_1} - {\hat w_1}{\phi _1} + {\dot \alpha _1})\\
\label{eq39}
&\dot{\hat{w}}_1=  - {e_2}{\phi _1}
\end{align}
where ${\hat w_1}{\phi _1}$ is the FLS to estimate nonlinearity. The Lyapunov function is selected and the derivation can be derived as follows:
\begin{align}
	\begin{split}\label{eq40}
		{V_e} =& \frac{1}{2}{e_1}^2 + \frac{1}{2}{e_2}^2 + \frac{1}{2}\tilde w_1^T{\tilde w_1}	
	\end{split}\\
	\begin{split}\label{eq41}
		{{\dot V}_e} =& {e_1}{{\dot e}_1} + {e_2}{{\dot e}_2} + \tilde w_1^T{{\dot \hat w}_1}\\
		=& {e_1}({e_2} + {\alpha _1}) + {e_2}(\theta \chi  + \frac{1}{M}(f + g)\\
		&- {{\dot \alpha }_1} + \theta u)+ \tilde w_1^T{{\dot \hat w}_1}\\
		=&  - {\lambda _1}e_1^2 - {\lambda _2}e_1^2	
    \end{split}
\end{align}
where ${\lambda _1}$, and ${\lambda _2}$ are positive constants. It is obvious that \eqref{eq41} is negative, which indicates that the system is asymptotic stable with fuzzy control law \eqref{eq38} and adaptive law \eqref{eq39}. Next, considering the use of PPF in the adaptive fuzzy controller, we transform the coordinate of the system state $[{x_{s1}},{x_{s2}}]$ as follow
\begin{align}
\begin{split}\label{eq42}
{\varepsilon _{s1}} =& \ln \left( {\frac{{a + \xi }}{{b - \xi }}} \right),(\xi  = \frac{{{x_{s1}}}}{\rho })
\end{split}\\
\begin{split}\label{eq43}
{\dot \varepsilon _{s1}} =& {\varepsilon _{s2}} = \frac{{a + b}}{{\left( {a + \xi } \right)\left( {b - \xi } \right)\rho }}({\dot x_{s1}} - {{{x_{s1}}\dot \rho } \mathord{\left/
		{\vphantom {{{x_{s1}}\dot \rho } \rho }} \right.
		\kern-\nulldelimiterspace} \rho })\\ 
	=& r({x_{s2}} - {{{x_{s1}}\dot \rho } \mathord{\left/
		{\vphantom {{{x_{s1}}\dot \rho } \rho }} \right.
		\kern-\nulldelimiterspace} \rho })
\end{split}\\
\begin{split}\label{eq44}
{{\dot \varepsilon }_{s2}} =& \dot r({x_{s2}} - {{{x_{s1}}\dot \rho } \mathord{\left/
			{\vphantom {{{x_{s1}}\dot \rho } \rho }} \right.
			\kern-\nulldelimiterspace} \rho }) + r\left( {{{\dot x}_{s2}} - \frac{{{x_{s2}}\dot \rho }}{\rho } - \frac{{{x_{s1}}\ddot \rho }}{\rho } + \frac{{{x_{s1}}{{\dot \rho }^2}}}{{{\rho ^2}}}} \right)\\
	=& \dot r({x_{s2}} - {{{x_{s1}}\dot \rho } \mathord{\left/
			{\vphantom {{{x_{s1}}\dot \rho } \rho }} \right.
			\kern-\nulldelimiterspace} \rho }) - r\left( {\frac{{{x_{s2}}\dot \rho }}{\rho } + \frac{{{x_{s1}}\ddot \rho }}{\rho } - \frac{{{x_{s1}}{{\dot \rho }^2}}}{{{\rho ^2}}}} \right) + r{{\dot x}_{s2}}\\
	=& F + r\theta u	
\end{split}
\end{align}
where $a,b > 0$ are  similarly defined as ${\delta _{Li}},{\delta _{Ui}}(i = 1,2,3)$ in Section 3.1, ${{r = \left( {a + b} \right)} \mathord{\left/
		{\vphantom {{r = \left( {a + b} \right)} {\left( {\left( {a + \xi } \right)\left( {b - \xi } \right)\rho } \right)}}} \right.
		\kern-\nulldelimiterspace} {\left( {\left( {a + \xi } \right)\left( {b - \xi } \right)\rho } \right)}}$ and $F = \dot r({x_{s2}} - {{{x_{s1}}\dot \rho } \mathord{\left/
		{\vphantom {{{x_{s1}}\dot \rho } \rho }} \right.
		\kern-\nulldelimiterspace} \rho }) - r\left( {{{{x_{s2}}\dot \rho } \mathord{\left/
			{\vphantom {{{x_{s2}}\dot \rho } \rho }} \right.
			\kern-\nulldelimiterspace} \rho } + {{{x_{s1}}\ddot \rho } \mathord{\left/
			{\vphantom {{{x_{s1}}\ddot \rho } \rho }} \right.
			\kern-\nulldelimiterspace} \rho } - {{{x_{s1}}{{\dot \rho }^2}} \mathord{\left/
			{\vphantom {{{x_{s1}}{{\dot \rho }^2}} {{\rho ^2}}}} \right.
			\kern-\nulldelimiterspace} {{\rho ^2}}}} \right) + r(\theta \chi  + {1 \mathord{\left/
		{\vphantom {1 M}} \right.
		\kern-\nulldelimiterspace} M}(f + g))$. $\rho $ is exponentially decreasing function defined by PPF. Moreover, one can define the transformed system error as ${s_1} = {\varepsilon _{s1}}$ and ${s_2} = {\varepsilon _{s2}} - {\alpha _2}$, where ${\alpha _2}$ is the intermediate virtual control and its expression is ${\alpha _2} =  - {\gamma _1}{s_1}$. Then with the help of the FLS, the adaptive fuzzy controller with PPF can be designed as follows:
\begin{align}
	\label{eq45} &u = \frac{1}{{r\theta }}( - {\gamma _2}{s_2} - {s_1} - {\hat w_2}{\phi _2} + {\dot \alpha _2})\\
	\label{eq46} &\dot {\hat{w}}_2 =  - {s_2}{\phi _2}
\end{align}	
where ${\hat w_2}{\phi _2}$ is the FLS for estimating the nonlinearity $F$. The Lyapunov function is defined and its derivation is obtained as follows:
\begin{align}	
\label{eq47}{V_s} =& \frac{1}{2}{s_1}^2 + \frac{1}{2}{s_2}^2 + \frac{1}{2}\tilde w_2^T{\tilde w_2}\\
\begin{split}\label{eq48}
	{{\dot V}_s} =& {s_1}{{\dot s}_1} + {s_2}{{\dot s}_2} + \tilde w_2^T{{\dot \hat w}_2}\\
	=& {s_1}({s_2} + {\alpha _2}) + {s_2}(F - {{\dot \alpha }_2} + r\theta u) + \tilde w_2^T{{\dot \hat w}_2}\\
	=&  - {\gamma _1}s_1^2 - {\gamma _2}s_2^2
\end{split}
\end{align}	
where ${\gamma _1}$, ${\gamma _2}$ are positive constant. Therefore ${\dot V_s}$ is negative and the transformed system is asymptotic stable. Then under control \eqref{eq45}, ${s_1}$ will converge to 0, thus $\xi  \to {{\left( {b - a} \right)} \mathord{\left/
		{\vphantom {{\left( {b - a} \right)} 2}} \right.
		\kern-\nulldelimiterspace} 2}$ and $\dot \xi  \to 0$. Besides, it should be noted that $\dot \xi  = {{\left( {{x_{s2}} - \xi \dot \rho } \right)} \mathord{\left/
		{\vphantom {{\left( {{x_{s2}} - \xi \dot \rho } \right)} \rho }} \right.
		\kern-\nulldelimiterspace} \rho }$, $\rho $ is the exponentially decreasing function and $\dot \rho $ is negative, thus the slope of $\dot \xi $ with respect to $\xi $ is positive. Consequently, when $\xi  > {{\left( {b - a} \right)} \mathord{\left/
		{\vphantom {{\left( {b - a} \right)} 2}} \right.
		\kern-\nulldelimiterspace} 2}$, then $\dot \xi  > 0$ and when $\xi  \le {{\left( {b - a} \right)} \mathord{\left/
		{\vphantom {{\left( {b - a} \right)} 2}} \right.
		\kern-\nulldelimiterspace} 2}$, then $\dot \xi  \le 0$. Based on the inequality \eqref{eq37} and discussion above, we have
\begin{equation}\label{eq49}
	\begin{split}
			&\left| {\ln \left( {\frac{{a + \xi }}{{b - \xi }}} \right) + \frac{{a + b}}{{\left( {a + \xi } \right)\left( {b - \xi } \right)}}\dot \xi } \right| \\
		&\ge \left| {\frac{4}{{a + b}}\left( {\xi  - \frac{{b - a}}{2}} \right) + \frac{4}{{a + b}}\dot \xi } \right|
	\end{split}
\end{equation}

Therefore, the salient convergence performance of adaptive fuzzy controller with PPF can be verified by comparing derivation \eqref{eq41} and \eqref{eq48}, that is ${\dot V_2} < {\dot V_1} \Rightarrow {\gamma _1}s_1^2 + {\gamma _2}s_1^2 > {k_1}e_1^2 + {k_2}e_1^2$, which is represented by $F(\xi ,\dot \xi )$  subsequently. One can derive the following equation using inequality \eqref{eq37} and \eqref{eq49}
\begin{equation}\label{eq50}
\begin{split}
	F(\xi &,\dot \xi )\\
	=& {\gamma _1}s_1^2 + {\gamma _2}s_2^2 - {\lambda _1}e_1^2 - {\lambda _2}e_2^2\\
	=& {\gamma _1}{\left( {\ln \left( {\frac{{a + \xi }}{{b - \xi }}} \right)} \right)^2} + {\gamma _2}\left( {\frac{{a + b}}{{\left( {a + \xi } \right)\left( {b - \xi } \right)}}\dot \xi } \right. + \\
	&{\left. {{\gamma _1}\ln \left( {\frac{{a + \xi }}{{b - \xi }}} \right)} \right)^2} - {\lambda _1}{\left( {\rho \xi } \right)^2} - {\lambda _2}{\left( {\rho \dot \xi  + ({k_1}\rho  + \dot \rho )\xi } \right)^2}\\
	\ge& {\gamma _1}{\left( {\frac{4}{{a + b}}\left( {\xi  - \frac{{b - a}}{2}} \right)} \right)^2} + {\gamma _2}\left( {{\gamma _1}\frac{4}{{a + b}}\left( {\xi  - \frac{{b - a}}{2}} \right)} \right.\\
	&+ {\left. {\frac{4}{{a + b}}\dot \xi } \right)^2} - {\lambda _1}{\left( {\rho \xi } \right)^2} - {\lambda _2}{\left( {({\lambda _1}\rho  + \dot \rho )\xi  + \rho \dot \xi } \right)^2}
\end{split}
\end{equation}
Substituting $X = \xi $, $Y = \dot \xi $, $c = {4 \mathord{\left/
		{\vphantom {4 {(a + b)}}} \right.
		\kern-\nulldelimiterspace} {(a + b)}}$, $d = {{(b - a)} \mathord{\left/
		{\vphantom {{(b - a)} 2}} \right.
		\kern-\nulldelimiterspace} 2}$, and $e = {k_1}\rho  + \dot \rho $, the above equation \eqref{eq50} can be further deduced as:
\begin{equation}\label{eq51}
\begin{split}
	F(X&,Y)\\
	\ge& {\gamma _1}{c^2}{\left( {X - d} \right)^2} + {\gamma _2}{\left( {{\gamma _1}c\left( {X - d} \right) + cy} \right)^2}\\
	&- {\lambda _1}{\rho ^2}{x^2} - {\lambda _2}{\left( {eX + \rho Y} \right)^2}\\
	=& {\gamma _1}{c^2}{\left( {X - d} \right)^2} + {\gamma _2}[{\gamma _1}^2{c^2}{(X - d)^2} + \\
	&2{\gamma _1}{c^2}(X - d)Y + {c^2}{Y^2}] - {\lambda _1}{\rho ^2}{X^2}\\
	&- {\lambda _2}[{e^2}{X^2} + 2e\rho XY + {\rho ^2}{Y^2}]\\
	=& ({\gamma _1}{c^2} - {\lambda _1}{\rho ^2}){\left( {X - \frac{{{\gamma _1}{c^2}d}}{{2({\gamma _1}{c^2} - {\lambda _1}{\rho ^2})}}} \right)^2}\\
	&+ ({\gamma _2}{c^2} - {\lambda _2}{\rho ^2}){\left( {Y + \frac{{{\gamma _1}{\gamma _2}{c^2} - {\lambda _2}e\rho }}{{{\gamma _2}{c^2} - {\lambda _2}{\rho ^2}}}X} \right)^2}\\
	&+ \left( {{\gamma _2}{\gamma _1}^2{c^2} - {\lambda _2}{e^2} - \frac{{{{({\gamma _1}{\gamma _2}{c^2} - {\lambda _2}e\rho )}^2}}}{{({\gamma _2}{c^2} - {\lambda _2}{\rho ^2})}}} \right){X^2}\\
	&- 2{\lambda _2}d{c^2}X - 2{\lambda _2}cdeY + {\lambda _2}{c^2}{d^2}\\
	&- \frac{{{{({\gamma _1}{c^2})}^2}{d^2}}}{{4({\gamma _1}{c^2} - {\lambda _1}{\rho ^2})}} + {\gamma _1}{c^2}{d^2}
\end{split}	
\end{equation}

The conditions to make \eqref{eq51} positive are that: 1) $({\gamma _1}{c^2} - {\lambda _1}{\rho ^2}) = {{{16{\gamma _1}} \mathord{\left/
			{\vphantom {{16{\gamma _1}} {(a + b)}}} \right.
			\kern-\nulldelimiterspace} {(a + b)}}^2} - {\lambda _1}{\rho ^2} \ge 0$; 2) $({\gamma _2}{c^2} - {\lambda _2}{\rho ^2}) = {{{16{\gamma _2}} \mathord{\left/
			{\vphantom {{16{\gamma _2}} {(a + b)}}} \right.
			\kern-\nulldelimiterspace} {(a + b)}}^2} - {\lambda _2}{\rho ^2} \ge 0$; 3) $\left( {{\gamma _2}{\gamma _1}^2{c^2} - {\lambda _2}{e^2} - {{{{({\gamma _1}{\gamma _2}{c^2} - {\lambda _2}e\rho )}^2}} \mathord{\left/
			{\vphantom {{{{({\gamma _1}{\gamma _2}{c^2} - {\lambda _2}e\rho )}^2}} {({\gamma _2}{c^2} - {\lambda _2}{\rho ^2})}}} \right.
			\kern-\nulldelimiterspace} {({\gamma _2}{c^2} - {\lambda _2}{\rho ^2})}}} \right) \ge 0$  and 4) $d = 0$ i.e., $a = b$. It is easy to meet the condition 4) because $a,b$ are user-defined. Note that $\rho $ is a function that decreases exponentially and eventually converges close to zero. Besides, ${{{16{\gamma _1}} \mathord{\left/
			{\vphantom {{16{\gamma _1}} {(a + b)}}} \right.
			\kern-\nulldelimiterspace} {(a + b)}}^2}$ and ${{{16{\gamma _2}} \mathord{\left/
			{\vphantom {{16{\gamma _2}} {(a + b)}}} \right.
			\kern-\nulldelimiterspace} {(a + b)}}^2}$ are positive constants. Consequently, conditions 1) and 2) always hold if ${{{16{\gamma _1}} \mathord{\left/
			{\vphantom {{16{\gamma _1}} {(a + b)}}} \right.
			\kern-\nulldelimiterspace} {(a + b)}}^2} \ge {\lambda _1}{\rho _0}^2$ and ${{{16{\gamma _2}} \mathord{\left/
			{\vphantom {{16{\gamma _2}} {(a + b)}}} \right.
			\kern-\nulldelimiterspace} {(a + b)}}^2} \ge {\lambda _2}{\rho _0}^2$, where ${\rho _0}$ is the initial value of $\rho $. As for condition 3), one can obtain that
		$$\begin{array}{l}
			\left( {{\gamma _2}{\gamma _1}^2{c^2} - {\lambda _2}{e^2} - \frac{{{{({\gamma _1}{\gamma _2}{c^2} - {\lambda _2}e\rho )}^2}}}{{({\gamma _2}{c^2} - {\lambda _2}{\rho ^2})}}} \right)\\
			= \frac{{16{{({\rho _1}({\gamma _1} - {\lambda _1}) + {{\dot \rho }_1})}^2}{\gamma _2}{\lambda _2}}}{{{{(a + b)}^2}{\lambda _2}{\rho ^2} - 16{\gamma _2}}},
		\end{array}$$
	and owing to the property of $\rho $, the condition 3) converges to zero. Also, one can set the parameter ${\gamma _1}$ close to ${\lambda _1}$.
	
In conclusion, the equation \eqref{eq51} is positive when the four above-mentioned conditions are satisfied; then the derivation \eqref{eq48} is smaller than \eqref{eq41}, which indicates that the convergence rate of the system under adaptive fuzzy control with PPF is greater than the one without.
	
$\boldsymbol{Remark 5}$ For the first time, this paper theoretically proves the superior convergence performance of PPF. The result can guide the design of the controller with PPF. It should be noted that the four conditions mentioned above are based on the internal properties of PPF parameters, such as $\rho $. Its exponential decrease characteristic determines the salient convergence performance. 

Though the convergence performance of the controller involved PPF is proved better, it still involves the FLS during the design. Thus, there is a need to discuss further the convergence property of the approximation-free controller involving PPF. Next, the convergence performance of the approximation-free controller proposed in this paper in comparison to the adaptive fuzzy controller \eqref{eq38} will be analyzed theoretically. According to the above analysis, it is ideal to set $a = b$, i.e., ${\delta _{Li}} = {\delta _{Ui}},i = 1,2,3$. From Section 3.2, one can obtain the derivation of Lyapunov function about relative suspension system under proposed control method as follows:
\begin{equation}\label{eq52}
\begin{split}
	{{\dot V}_\varepsilon } =& {{\dot V}_1} + {{\dot V}_2} + {{\dot V}_3}\\
	\le&  - {r_1}({k_1}{\varepsilon _1}^2 - {Q_1}\left| {{\varepsilon _1}} \right|) - {r_2}({k_2}{\varepsilon _2}^2 - {Q_2}\left| {{\varepsilon _2}} \right|)\\
	&- {r_3}({k_3}{\varepsilon _3}^2 - {Q_3}\left| {{\varepsilon _3}} \right|)
\end{split}
\end{equation}	

Based on the knowledge in Section 3.2, one can get that ${\xi _3} = {{\left( {{x_2} - {u_2}} \right)} \mathord{\left/
		{\vphantom {{\left( {{x_2} - {u_2}} \right)} {{\rho _3}}}} \right.
		\kern-\nulldelimiterspace} {{\rho _3}}}$, $({u_2} =  - \left( {{{{k_2}} \mathord{\left/
			{\vphantom {{{k_2}} \theta }} \right.
			\kern-\nulldelimiterspace} \theta }} \right){\varepsilon _2})$ and ${x_2} = {\xi _2}{\rho _2} + {u_1} + {y_2} + {x_4},({u_1} =  - {k_1}{\varepsilon _1})$. By virtue of inequality \eqref{eq37} and $AB \le {{\left( {{A^2} + {B^2}} \right)} \mathord{\left/
		{\vphantom {{\left( {{A^2} + {B^2}} \right)} 2}} \right.
	\kern-\nulldelimiterspace} 2}$, the third term on the right side of \eqref{eq52} is further expressed as:
\begin{equation}\label{eq53}
\begin{split}
	{r_3}({k_3}{\varepsilon _3}^2& - {Q_3}\left| {{\varepsilon _3}} \right|)\\
	\ge& 2{r_3}\left( {\frac{{2{k_3} - 1}}{{{a^2}}}} \right){\xi _3}^2 - \frac{1}{2}{r_3}{Q_3}^2\\
	=& \beta \left( {{x_2}^2 - 2{x_2}{u_2} + {u_2}^2} \right) - \frac{1}{2}{r_3}{Q_3}^2\\
	=& \beta {u_2}^2 + \beta {u_1}^2 + 2\beta \left( {{\xi _2}{\rho _2} + {y_2} + {x_4}} \right)\\
	&+ \beta {\left( {{\xi _2}{\rho _2} + {y_2} + {x_4}} \right)^2} - 2\beta {x_2}{u_2} - \frac{1}{2}{r_3}{Q_3}^2\\
	=& \beta {{{({k_2}} \mathord{\left/
				{\vphantom {{({k_2}} {\theta )}}} \right.
				\kern-\nulldelimiterspace} {\theta )}}^2}{\varepsilon _2}^2 + \beta {k_1}^2{\varepsilon _1}^2 + 2\beta \Delta {u_1}\\
	&+ \beta {\Delta ^2} - 2\beta {x_2}{u_2} - \frac{1}{2}{r_3}{Q_3}^2
\end{split}
\end{equation}
wherein, $\beta  = {{2{r_3}\left( {2{k_3} - 1} \right)} \mathord{\left/
		{\vphantom {{2{r_3}\left( {2{k_3} - 1} \right)} {\left( {{a^2}{\rho _3}^2} \right)}}} \right.
		\kern-\nulldelimiterspace} {\left( {{a^2}{\rho _3}^2} \right)}}$ and $\Delta  = {\xi _2}{\rho _2} + {y_2} + {x_4}$. Then subtracting \eqref{eq41} from \eqref{eq52}, substituting \eqref{eq53}, and utilizing inequality \eqref{eq37}, the following equation is derived
\begin{align}\label{54}
	\left| {{{\dot V}_\varepsilon }} \right|& - \left| {{{\dot V}_z}} \right| \notag \\
	=& {r_1}({k_1}{\varepsilon _1}^2 - {{\bar Q}_1}\left| {{\varepsilon _1}} \right|) + {r_2}({k_2}{\varepsilon _2}^2 - {{\bar Q}_2}\left| {{\varepsilon _2}} \right|) \notag \\
	&+ {r_3}({k_3}{\varepsilon _3}^2 - {{\bar Q}_3}\left| {{\varepsilon _3}} \right|) - {\lambda _1}{e_1}^2 - {\lambda _2}{e_2}^2 \notag \notag \\
	\ge& {r_1}\left( {\frac{{4{k_1}}}{{{a^2}}}{\xi _1}^2 - {{\bar Q}_1}\left| {{\varepsilon _1}} \right|} \right) + {r_2}\left( {\frac{{4{k_2}}}{{{a^2}}}{\xi _2}^2 - {{\bar Q}_2}\left| {{\varepsilon _2}} \right|} \right) \notag \\
	&- {\lambda _1}{\rho _1}^2{\xi _1}^2 - {\lambda _2}{({\lambda _1}{\rho _1}{\xi _1} - {k_1}{\varepsilon _1} + {\rho _2}{\xi _2})^2} \notag \\
	&+ {r_3}({k_3}{\varepsilon _3}^2 - {{\bar Q}_3}\left| {{\varepsilon _3}} \right|) \notag \\
	=& \left( {\frac{{4{k_1}{r_1}}}{{{a^2}}} - {\lambda _1}^2{\lambda _2}{\rho _1}^2 - {\lambda _1}{\rho _1}^2} \right){\xi _1}^2 - 2{\lambda _1}{\lambda _2}{\rho _1}{\rho _2}{\xi _1}{\xi _2} \notag \\
	&+ \left( {\frac{{4{k_2}{r_2}}}{{{a^2}}} - {\lambda _2}{\rho _2}^2} \right){\xi _2}^2 - {\lambda _2}{k_1}^2{\varepsilon _1}^2 \notag \\
	&+ (2{k_1}{\lambda _1}{\lambda _2}{\rho _1}{\xi _1} + 2{k_1}{\lambda _2}{\rho _2}{\xi _2}){\varepsilon _1} - {r_1}{{\bar Q}_1}\left| {{\varepsilon _1}} \right| \notag \\
	&- {r_2}{{\bar Q}_2}\left| {{\varepsilon _2}} \right| + {r_3}({k_3}{\varepsilon _3}^2 - {{\bar Q}_3}\left| {{\varepsilon _3}} \right|) \notag \\
	=& \left( {\frac{{4{k_1}{r_1}}}{{{a^2}}} - {\lambda _1}^2{\lambda _2}{\rho _1}^2 - {\lambda _1}{\rho _1}^2} \right){\left( {{\xi _1} - \eta {\xi _2}} \right)^2} \notag \\
	&+ \left( {\frac{{4{k_2}{r_2}}}{{{a^2}}} - {\lambda _2}{\rho _2}^2 - \mu } \right){\xi _2}^2 + \left( {{\lambda _2}{k_1}^2 + \beta {k_1}^2} \right){\varepsilon _1}^2 \notag \\
	&- \left( {2\beta {k_1}\bar \Delta  + {r_1}{{\bar Q}_1}} \right)\left| {{\varepsilon _1}} \right| + \beta {{{({k_2}} \mathord{\left/
				{\vphantom {{({k_2}} {\theta )}}} \right.
				\kern-\nulldelimiterspace} {\theta )}}^2}{\varepsilon _2}^2 \notag \\
	&- \left( {2\beta {{({k_2}} \mathord{\left/
				{\vphantom {{({k_2}} {\theta )}}} \right.
				\kern-\nulldelimiterspace} {\theta )}}{{\bar x}_2} + {r_2}{{\bar Q}_2}} \right)\left| {{\varepsilon _2}} \right| + \beta {\Delta ^2} - \frac{1}{2}{r_3}{{\bar Q}_3}^2
\end{align}	
wherein, $\bar \Delta  = \sup \left\{ {\left| \Delta  \right|} \right\}$, ${\bar x_2} = \sup \left\{ {\left| {{x_2}} \right|} \right\}$, $\eta  = {\lambda _1}{\lambda _2}{\rho _1}{\rho _2}{a^2}/(4{r_1}{k_1} - {a^2}{\lambda _1}^2{\lambda _2}{\rho _1}^2 - {a^2}{\lambda _1}{\rho _1}^2)$ and $\mu  = {{{\lambda _1}^2{\lambda _2}^2{\rho _1}^2{\rho _2}^2{a^2}} \mathord{\left/{\vphantom {{{\lambda _1}^2{\lambda _2}^2{\rho _1}^2{\rho _2}^2{a^2}} {\left( {4{r_1}{k_1} - {a^2}{\lambda _1}^2{\lambda _2}{\rho _1}^2 - {a^2}{\lambda _1}{\rho _1}^2} \right)}}} \right.\kern-\nulldelimiterspace} {\left( {4{r_1}{k_1} - {a^2}{\lambda _1}^2{\lambda _2}{\rho _1}^2 - {a^2}{\lambda _1}{\rho _1}^2} \right)}}$. Thus, from the equation \eqref{54}, it can be inferred that the conditions ${{4{r_1}{k_1}} \mathord{\left/{\vphantom {{4{r_1}{k_1}} {{a^2}}}} \right.\kern-\nulldelimiterspace} {{a^2}}} - {\lambda _1}^2{\lambda _2}{\rho _1}^2 - {\lambda _1}{\rho _1}^2 \ge 0$ and ${{4{r_2}{k_2}} \mathord{\left/	{\vphantom {{4{r_2}{k_2}} {{a^2}}}} \right.\kern-\nulldelimiterspace} {{a^2}}} - {\lambda _2}{\rho _2}^2 - \mu  \ge 0$ are always satisfied if they are satisfied at initial position, i.e., at ${r_1} = {r_{10}},{r_2} = {r_{20}}, {\rho _1} = {\rho _{10}},{\rho _2} = {\rho _{20}}$, owing to the exponentially decrease property of ${\rho _1}$ and ${\rho _2}$. In addition, it is obvious that ${\lambda _2}{k_1}^2 + \beta {k_1}^2$ and $\beta {{{({k_2}} \mathord{\left/{\vphantom {{({k_2}} {\theta )}}} \right.\kern-\nulldelimiterspace} {\theta )}}^2}$ in \eqref{54} are positive. Then, with the knowledge that ${\varepsilon _1} > {{{{\bar Q}_1}} \mathord{\left/{\vphantom {{{{\bar Q}_1}} {{k_1}}}} \right.\kern-\nulldelimiterspace} {{k_1}}}$, ${\varepsilon _2} > {{{{\bar Q}_2}} \mathord{\left/{\vphantom {{{{\bar Q}_2}} {{k_2}}}} \right.\kern-\nulldelimiterspace} {{k_2}}}$ and ${\varepsilon _3} > {{{{\bar Q}_3}} \mathord{\left/	{\vphantom {{{{\bar Q}_3}} {{k_3}}}} \right.\kern-\nulldelimiterspace} {{k_3}}}$ provided in Section 3.2, one can deduce the following conditions
\begin{align*}
	\frac{{\left( {2\beta \bar \Delta {k_1} + {r_1}{{\bar Q}_1}} \right)}}{{\left( {{\lambda _2} + \beta } \right){k_1}^2}} \le \frac{{{{\bar Q}_1}}}{{{k_1}}} \Rightarrow \frac{{2\beta {k_1}\bar \Delta }}{{\left( {{\lambda _2} + \beta } \right){k_1} - {r_1}}} \le {\bar Q_1}\\
	\frac{{\left( {2\beta {{({k_2}} \mathord{\left/
						{\vphantom {{({k_2}} {\theta )}}} \right.
						\kern-\nulldelimiterspace} {\theta )}}{{\bar x}_2} + {r_2}{{\bar Q}_2}} \right)}}{{\beta {{{{({k_2}} \mathord{\left/
							{\vphantom {{({k_2}} {\theta )}}} \right.
							\kern-\nulldelimiterspace} {\theta )}}}^2}}} \le \frac{{{{\bar Q}_2}}}{{{k_2}}} \Rightarrow \frac{{2\beta {{({k_2}} \mathord{\left/
					{\vphantom {{({k_2}} {\theta )}}} \right.
					\kern-\nulldelimiterspace} {\theta )}}{{\bar x}_2}}}{{\beta {{({k_2}} \mathord{\left/
					{\vphantom {{({k_2}} {{\theta ^2})}}} \right.
					\kern-\nulldelimiterspace} {{\theta ^2})}} - {r_2}}} \le {\bar Q_2}.
\end{align*}

Furthermore, the last term of \eqref{54} indicates that if $\beta {\Delta ^2} - {{{r_3}{{\bar Q}_3}^2} \mathord{\left/
		{\vphantom {{{r_3}{{\bar Q}_3}^2} 2}} \right.
		\kern-\nulldelimiterspace} 2} \ge 0$ is satisfied, $\beta $ needs to be appropriately chosen a large value. The convergence performance analysis above helps to choose proper parameters in subsequent simulation.

\section{Simulation and discussion}
Specific examples are presented to verify the feasibility and superiority of the proposed controller and the comparison between the proposed controller and another controller is executed here. In order to facilitate subsequent simulation and comparison, the unknown nonlinear terms in the suspension system \eqref{eq1} are given as follow
\begin{align}\label{eq55}
	{F_s} &= {k_{s1}}({z_s} - {z_u}) + {k_{s2}}{({z_s} - {z_u})^2} + {k_{s3}}{({z_s} - {z_u})^3}\notag\\
	{F_d} &= {c_{s1}}({{\dot z}_s} - {{\dot z}_u}) + {c_{s2}}{({{\dot z}_s} - {{\dot z}_u})^2}
\end{align}
wherein ${k_{s1}}$ is linear stiffness coefficient; ${k_{s2}}$ and ${k_{s3}}$ are nonlinear stiffness coefficients; ${c_{s1}}$ and ${c_{s2}}$ are damping coefficients of nonlinear dampers; ${k_t}$ and ${c_t}$ denote stiffness coefficient and damping coefficient of the tire, respectively.

Moreover, to verify the superiority, the fuzzy adaptive control (FAC) based on the bioinspired model for a suspension system in [31] is compared to the controller proposed in this paper. The parameters of the suspension system and bioinspired reference model are given in Table \ref{table_1} and Table \ref{table_2}. 

\begin{table}
	\begin{center}
		\caption{Parameters of suspension system \cite{b31}}
		\label{table_1}
		\setlength{\tabcolsep}{3pt}
		\begin{tabular}{|p{45pt}|p{67.5pt}|p{45pt}|p{67.5pt}|}
			\hline Parameter & Value        & Parameter     & Value      \\
			\hline
			${m_s}$	        & 240kg        & ${k_t}$            & 23.61kg    \\
			${m_u}$	        & 15394N/m     & ${c_t}$            & 1385.4Ns/m \\
			${k_{s1}}$  	& -73696N/m    & ${c_{s1}}$         & 524.28Ns/m \\
			${k_{s2}}$	    & 3170400N/m   & ${c_{s2}}$         & 13.8Ns/m   \\
			${k_{s3}}$   	& 181818.88N/m &                    &            \\
			\hline
		\end{tabular}
	\end{center}	
\end{table}

\begin{table}
	\begin{center}
		\caption{Parameters of bioinspired model \cite{b31}}
		\label{table_2}
		\setlength{\tabcolsep}{3pt}
		\begin{tabular}{|p{45pt}|p{67.5pt}|p{45pt}|p{67.5pt}|}
			\hline Parameter & Value        & Parameter     & Value      \\
			\hline
			$M$	            & 240kg                     & ${k_v}$            & 250N/m    \\
			${L_1}$	        & 0.1m                      & ${k_h}$            & 500N/m \\
			${L_2}$  	    & 0.2m                      & ${\mu _1}$         & 1Ns/m \\
			${\theta _1}$   & ${\pi  \mathord{\left/
					{\vphantom {\pi  6}} \right.
					\kern-\nulldelimiterspace} 6}$ rad  & ${\mu _2}$         & 0.155Ns/m   \\	
			\hline
		\end{tabular}
	\end{center}	
\end{table}

Three types of road excitation were used here: random road, bump road, and sinusoidal road. The random road was used to verify the vibration suppression performance of the proposed control method, and bump road and sinusoidal road were utilized to show the transient performance.

\subsection{Random Road}

\begin{table}[]
	\begin{center}
		\caption{Acceleration RMS of suspension systems with different controllers (m/s2)}
		\label{table_3}
		\setlength{\tabcolsep}{3pt}
		\begin{tabular}{|p{65pt}|p{30pt}|p{65pt}|p{65pt}|}
			\hline
			V & Passive & FAC{[}31{]}      & Proposed         \\
			\hline
			${V_1} = 20km/h$		& 0.116   & 0.0292(↓74.82\%) & 0.0240(↓79.31\%) \\
			${V_2} = 40km/h$		& 0.164   & 0.0413(↓74.82\%) & 0.0342(↓79.15\%) \\
			${V_3} = 60km/h$		& 0.2009  & 0.0506(↓74.81\%) & 0.0419(↓79.14\%) \\
			${V_4} = 100km/h$		& 0.2593  & 0.0655(↓74.74\%) & 0.0542(↓79.10\%) \\
			\hline
		\end{tabular}
	\end{center}	
\end{table}

\begin{table}[]
	\begin{center}
		\caption{Computational time (s)}
		\label{table_4}
		\setlength{\tabcolsep}{3pt}
		\begin{tabular}{|p{65pt}|p{65pt}|p{95pt}|}
			\hline
			V & FAC{[}31{]} & Proposed        \\
			\hline
           ${V_1} = 20km/h$		& 204.5893    & 44.5253(↓78.24\%) \\
           ${V_2} = 40km/h$		& 204.4992    & 45.7517(↓77.63\%) \\
           ${V_3} = 60km/h$		& 205.3395    & 44.8423(↓78.16\%) \\
           ${V_4} = 100km/h$	& 204.3396    & 45.9872(↓77.49\%) \\
			\hline
		\end{tabular}
	\end{center}	
\end{table}

\begin{table}[]
	\begin{center}
		\caption{Acceleration RMS under different controllers with disturbance (m/s2)}
		\label{table_5}
		\setlength{\tabcolsep}{3pt}
		\begin{tabular}{|p{65pt}|p{30pt}|p{65pt}|p{65pt}|}
			\hline
			V & Passive & FAC{[}31{]}      & Proposed         \\
			\hline
			${V_1} = 20km/h$	&1.225	&0.0662(↓94.60\%)	&0.0454(↓96.29\%)\\
			${V_2} = 40km/h$	&1.2306	&0.0745(↓93.95\%)	&0.0522(↓95.76\%)\\
			${V_3} = 60km/h$	&1.2363	&0.0819(↓93.38\%)	&0.0582(↓95.29\%) \\
			${V_4} = 100km/h$	&1.2478	&0.0952(↓92.37\%)	&0.0692(↓94.45\%) \\
			\hline
		\end{tabular}
	\end{center}	
\end{table}

\begin{table}[]
	\begin{center}
		\caption{Computational time with disturbance (s)}
		\label{table_6}
		\setlength{\tabcolsep}{3pt}
		\begin{tabular}{|p{65pt}|p{65pt}|p{95pt}|}
			\hline
			V & FAC{[}31{]} & Proposed        \\
			\hline
			${V_1} = 20km/h$		&337.7891	&53.6581(↓84.11\%) \\
			${V_2} = 40km/h$		&340.0640	&51.5062(↓84.85\%) \\
			${V_3} = 60km/h$		&338.2901	&50.8201(↓84.98\%) \\
			${V_4} = 100km/h$	    &341.2300	&52.7526(↓84.54\%) \\
			\hline
		\end{tabular}
	\end{center}	
\end{table}

The random excitation is present as follow
\begin{equation}\label{eq56}
	{\dot z_r}(t) =  - 2\pi {n_z}V{z_r}(t) + 2\pi {n_0}\sqrt {{G_z}({n_0})V} \omega (t)
\end{equation}
where ${n_z} = 0.0001$, ${n_0} = 0.1$ and ${G_z}({n_0}) = 256 \times {10^{ - 6}}$; $\omega (t)$ is the standard Gaussian white noise with 0 mean and unit variance and $V$ is the longitudinal velocity of the vehicle. Four different velocities were set, i.e., $V = 20km/h$, $V = 40km/h$, $V = 60km/h$, and $V = 100km/h$ for subsequent simulation. The whole simulation time was taken as 50s.

The initial states of the system were all set as zero. Parameters of the proposed controller were chosen through the trial-and-error method. Hence, control gains were set as ${k_1} = 0.01$, ${k_2} = 0.083$ and ${k_3} = 0.834$. Parameters of tanh function were defined as ${\delta _{L1}} = {\delta _{U1}} = {\delta _{L2}} = {\delta _{U2}} = {\delta _{L3}} = {\delta _{U3}} = 1$ and prescribed performance functions were given as follows:
$${\rho _1} = (0.001 - 0.0001){e^{ - 17t}} + 0.0001$$
$${\rho _2} = (0.55 - 0.1){e^{ - 15t}} + 0.1$$
$${\rho _3} = (1.1 - 0.95){e^{ - 12t}} + 0.95$$

To evaluate the vibration isolation performance of different controllers, the root-mean-square (RMS) was used as the evaluation criteria. The general form can be described as follows:
\begin{equation}\label{eq57}
	RMS(\chi ) = \sqrt {\frac{1}{T}\int_0^T {{\chi ^2}dt} }
\end{equation}

\begin{figure}[!t]
	\centerline{\includegraphics[width=\columnwidth]{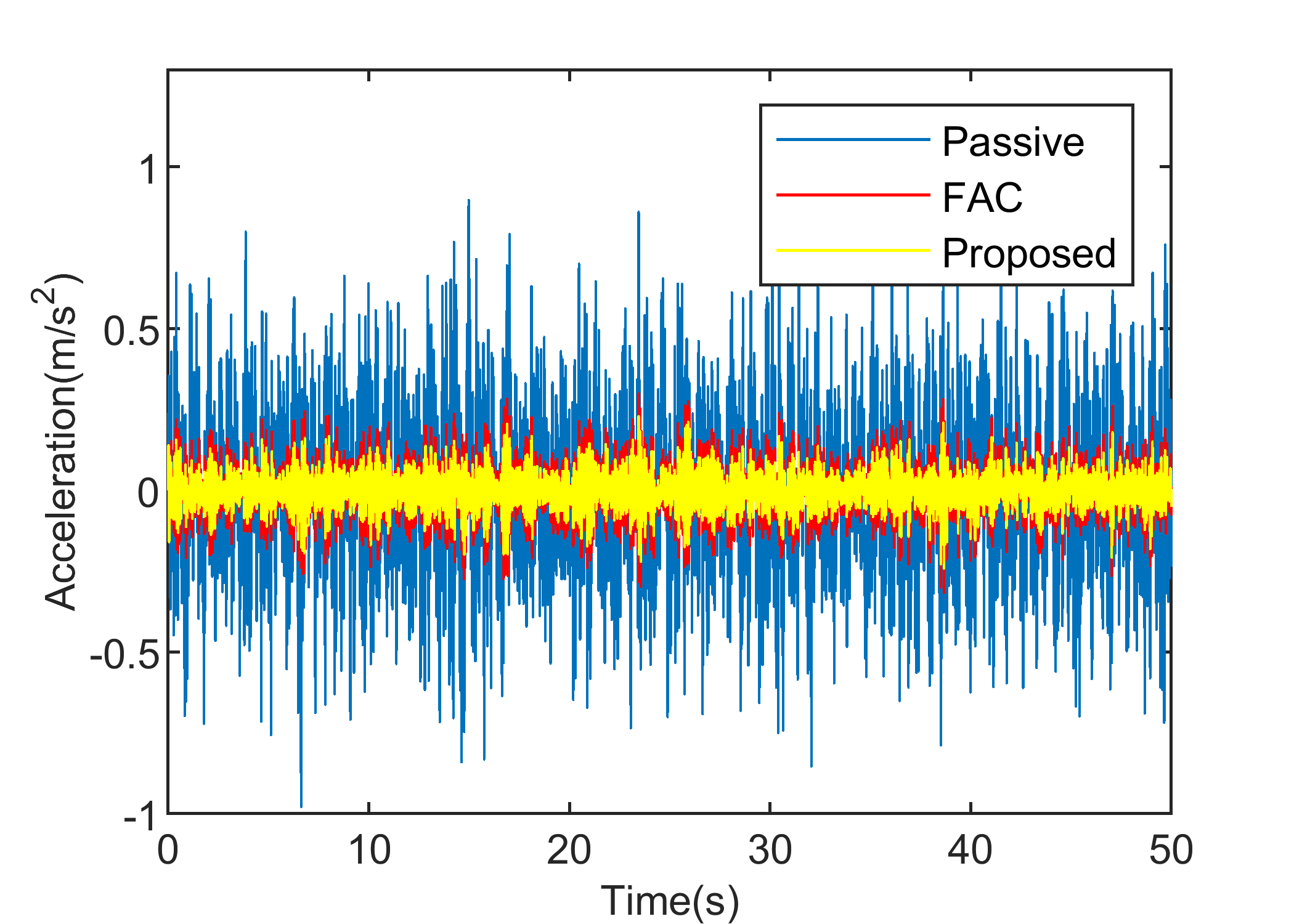}}
	\caption{Acceleration of suspension systems with different controllers for $V = 100km/h$}
	\label{fig_4}
\end{figure}

 Only the time history diagram of sprung mass acceleration with different controllers when $V = 100km/h$ is shown in Fig \ref{fig_4} as the representative (similar trends were observed for the other three groups of velocities) and corresponding acceleration RMSs of suspension system controlled by different controllers in different velocities are listed in Table \ref{table_3}, where the passive suspension system is the reference. In Fig \ref{fig_4}, the blue, red and yellow lines represent the acceleration produced by the passive suspension system, suspension with FAC, and suspension with the proposed controller, respectively. From Fig \ref{fig_4}, it can be concluded that both two controllers effectively suppressed the vibration compared to the passive suspension system and the acceleration of the suspension system with the proposed control method decreased more than that could be achieved with the FAC, which can be further verified in Table \ref{table_3}. The proposed controller reduces the RMS of acceleration more than FAC even when velocity varies, which indicates better vibration suppression. Besides, the actual time taken by the  suspension systems with different controllers is given in Table \ref{table_4}. Compared to the suspension system with FAC, the one with the proposed control method significantly reduces the computational time, which illustrates the simplicity of the proposed controller.

\begin{figure}[!t]
	\subfigure[Dynamic tyre load]
	{\centering\includegraphics[width=0.5\columnwidth]{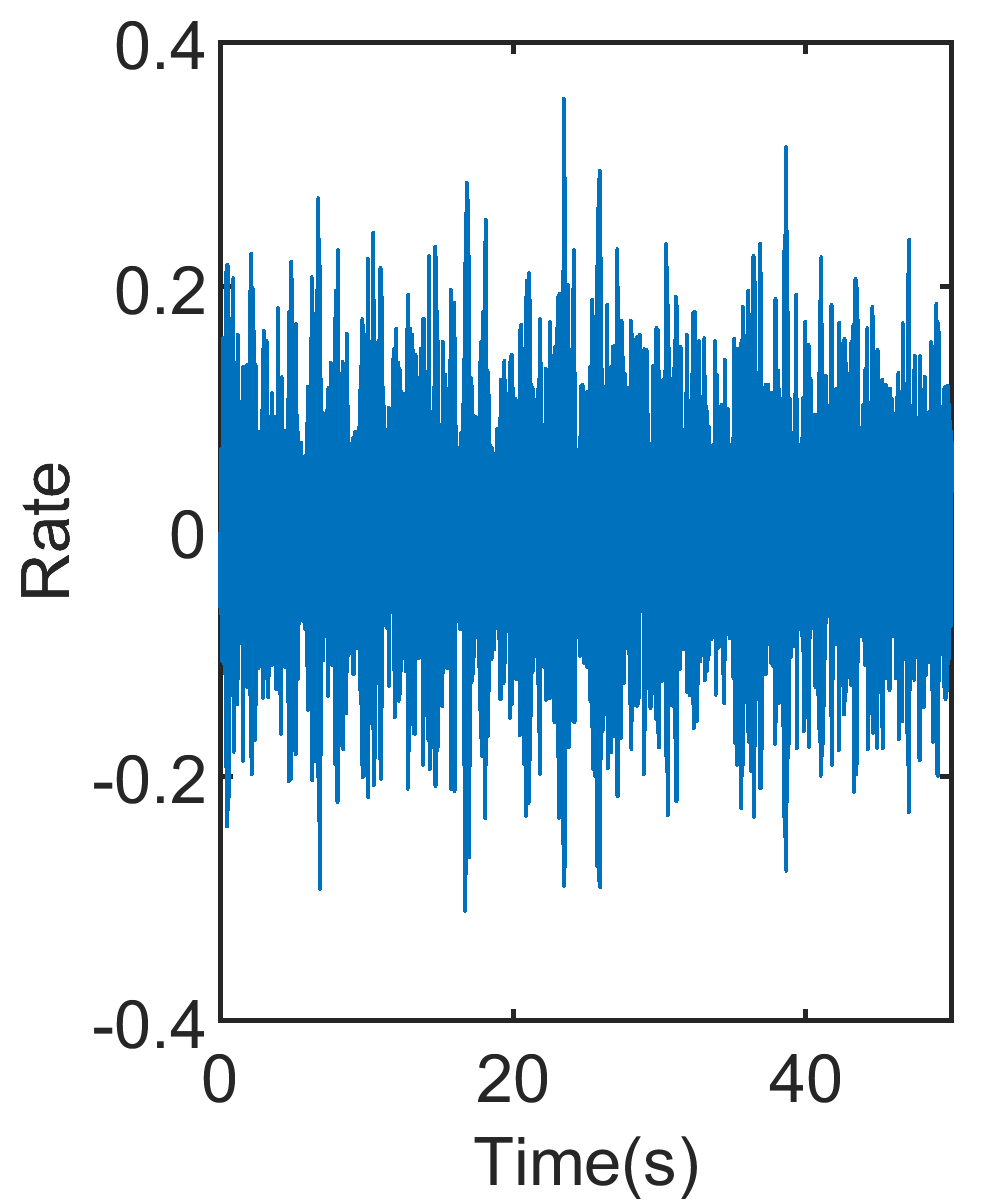}}%
	\subfigure[Suspension deflection]
	{\centering\includegraphics[width=0.5\columnwidth]{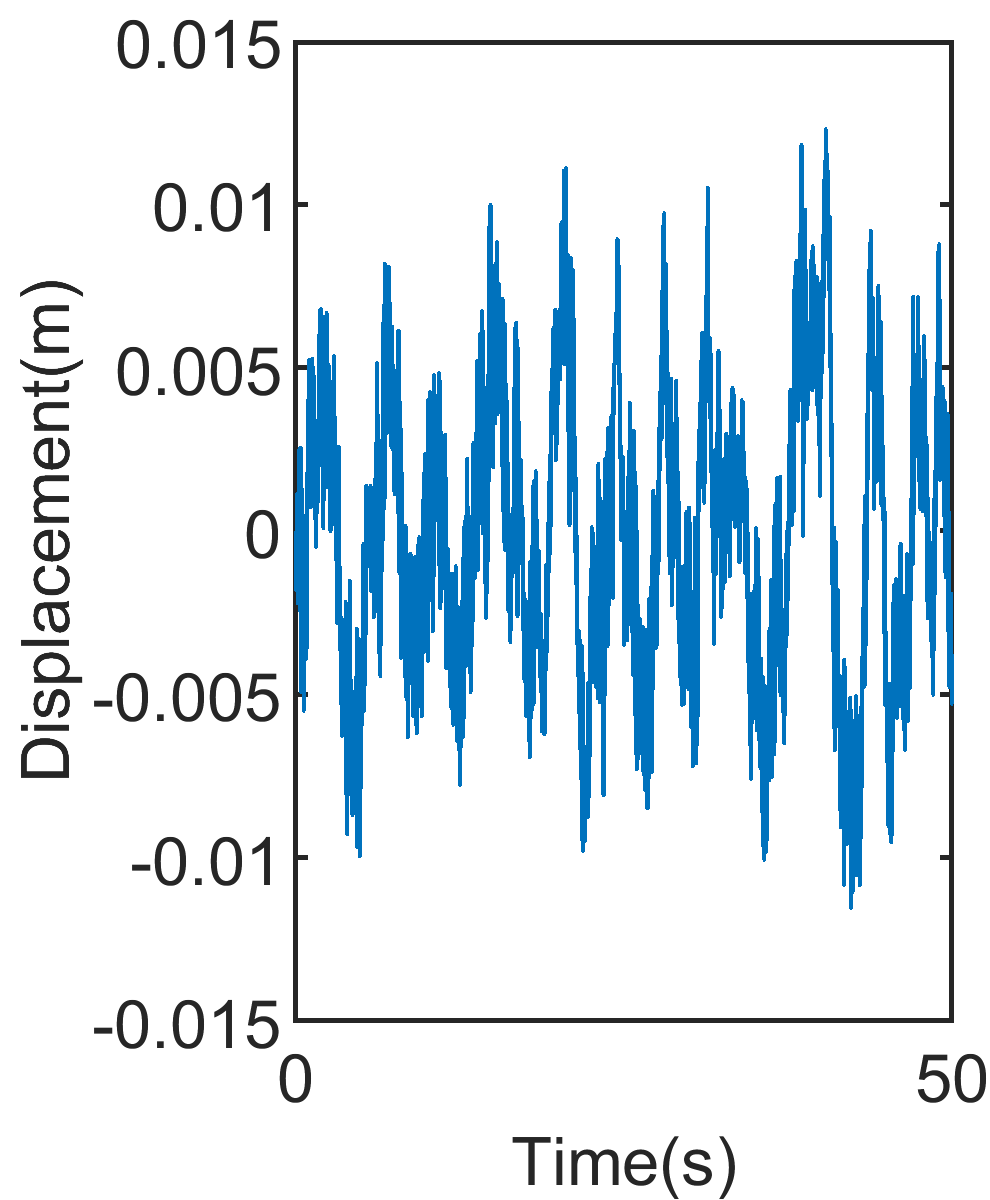}}
	\caption{Acceleration of suspension systems under different controllers when $V = 100km/h$}
	\label{fig_5}
\end{figure}

As safety and comfort improvement must be guaranteed, the suspension system should satisfy the constraints \eqref{eq4} and \eqref{eq5} metioned above in Section 2. The situation during the maximum velocity $V = 100km/h$ is analyzed representatively here. The maximum suspension deflection is ${z_{\max }} = 0.1m$. The dynamic tyre load and suspension deflection are shown in Fig \ref{fig_5} (a, b), wherein the rate represents the dynamic tyre load ${{\left( {{F_t} + {F_b}} \right)} \mathord{\left/
		{\vphantom {{\left( {{F_t} + {F_b}} \right)} {({m_s} + {m_u})g}}} \right.
		\kern-\nulldelimiterspace} {({m_s} + {m_u})g}}$. From Fig \ref{fig_5}, it is clear that both indexes are in the acceptable range, proving that the proposed controller improve the suspension system’s comfort and ensures safety simultaneously.

In order to demonstrate the robustness of the proposed controller, the disturbance $w = \sin (3\pi t) + 0.2\sin (30\pi t)$ is introduced into sprung mass in \eqref{eq1}. The same random road excitation is used, and the same four forward velocities are chosen, then the comparison of acceleration RMS and computational time under different controllers are presented in Table \ref{table_5} and Table \ref{table_6}, respectively, when disturbance exists. FAC and the proposed controller can resist disturbance, as shown in Table \ref{eq5}. Furthermore, the proposed control method performs better than FAC in suppressing the vibration during various velocities when disturbance exists, which illustrates the better robustness of proposed controller. From Table \ref{table_6}, the conclusion can be made that the proposed controller’s computational time is still less than FAC’s. Besides, the reduction in computational time with disturbance situation is more than the reduction in computational time without disturbance, which further indicates the superior robustness of the proposed controller.

\begin{figure}[!t]
	\centerline{\includegraphics[width=\columnwidth]{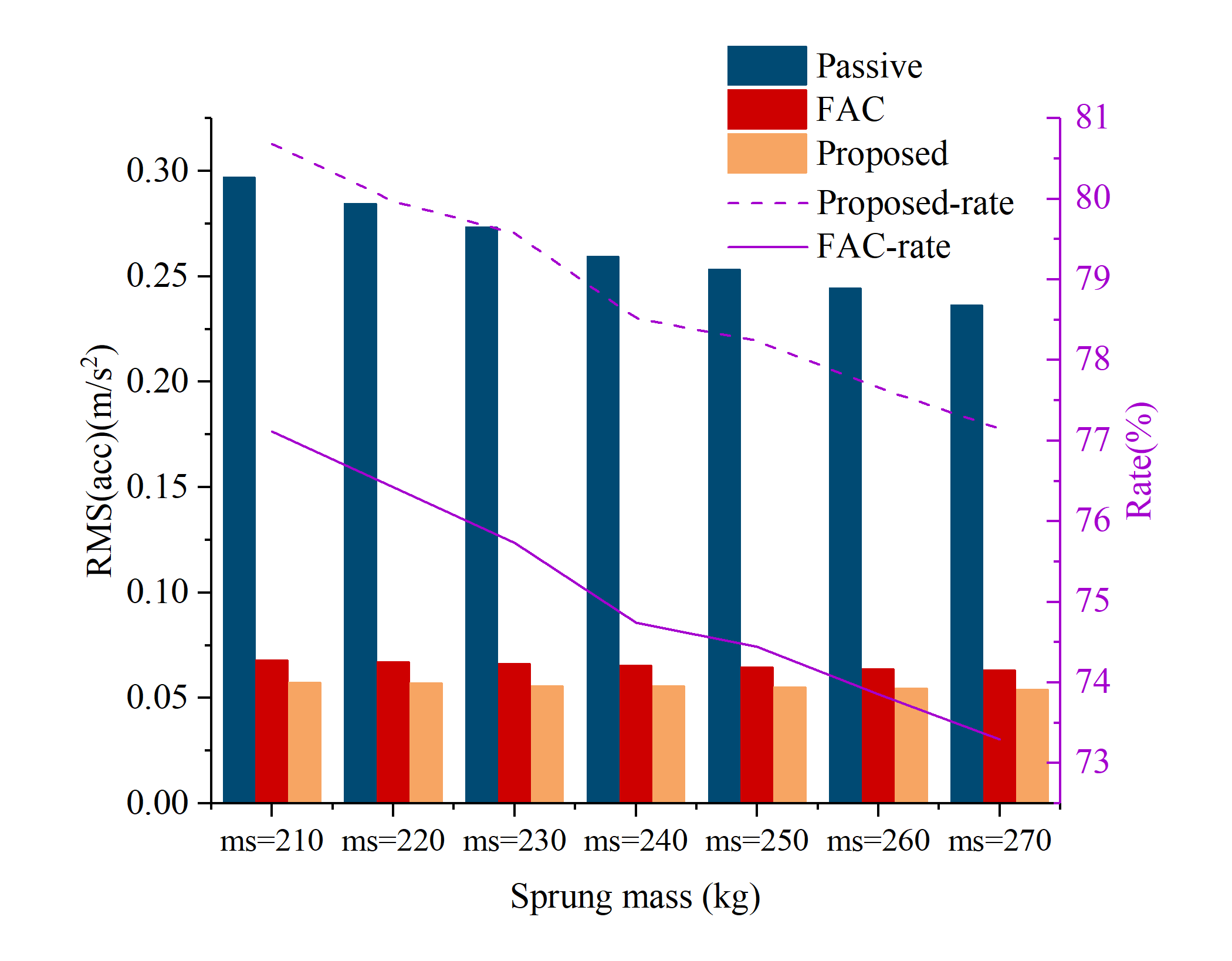}}
	\caption{RMS of acceleration with sprung mass changing}
	\label{fig_6}
\end{figure}

In addition, to testify the adaptation of the proposed control method to uncertainty, the simulation is executed when the sprung mass varies from 210kg to 270kg in the interval of 10 kg. Using the same random road with the forward velocity $V = 100km/h$, the result is shown in Fig \ref{fig_6}, wherein the Proposed-rate and FAC-rate express the ratio of the acceleration RMS under the proposed scheme and FAC respected to the one of passive suspension, respectively. It is evident that all acceleration RMSs with different suspension system settings (Proposed, FAC, Passive) decrease with the increase of sprung mass, but the proposed controller still performs better in suppressing vibration compared to FAC, implying that the proposed controller is more adaptive to uncertainty.


\subsection{Bump Road}
 The simulation is implemented under bump road excitation to verify the convergence performance of the proposed controller. The following equation expresses the bump road.
\begin{equation}\label{eq58}
{z_r}(t) = \left\{ \begin{split}{l}
	\frac {\alpha }{4}(1 - &\cos (\frac{{2\pi V}}{l}t)),{\rm{  }} & 0 \le t \le \frac{l}{V}\\
	&0, & t > \frac{l}{V}
\end{split} \right.
\end{equation}
where $a$ , $l$ are the height and length of the bump, which is set as $\alpha  = 0.1m,l = 5m$, and $V$ is the forward velocity, which is chosen as $V = 40km/h$. The initial states of the system are set as  $[{x_1},{x_2},{x_3},{x_4}] = [0.06,0,0,0]$, and $[{y_1},{y_2}] = [0,0]$. The control parameters are given as follows: ${k_1} = 3$, ${k_2} = 0.033$, ${k_3} = 4.167$ and tanh function parameters are selected as ${\delta _{L1}} = {\delta _{U1}} = {\delta _{L2}} = {\delta _{U2}} = {\delta _{L3}} = {\delta _{U3}} = 1$. Prescribed performance functions are determined as
$${\rho _{U1}} = (0.1 - 0.0001){e^{ - 8t}} + 0.0001,{\rho _{L1}} = 0.0001$$
$${\rho _2} = (1 - 0.01){e^{ - 5t}} + 0.01$$
$${\rho _3} = (12.5 - 0.5){e^{ - 1t}} + 0.5.$$

\begin{figure}[!t]
	\centerline{\includegraphics[width=\columnwidth]{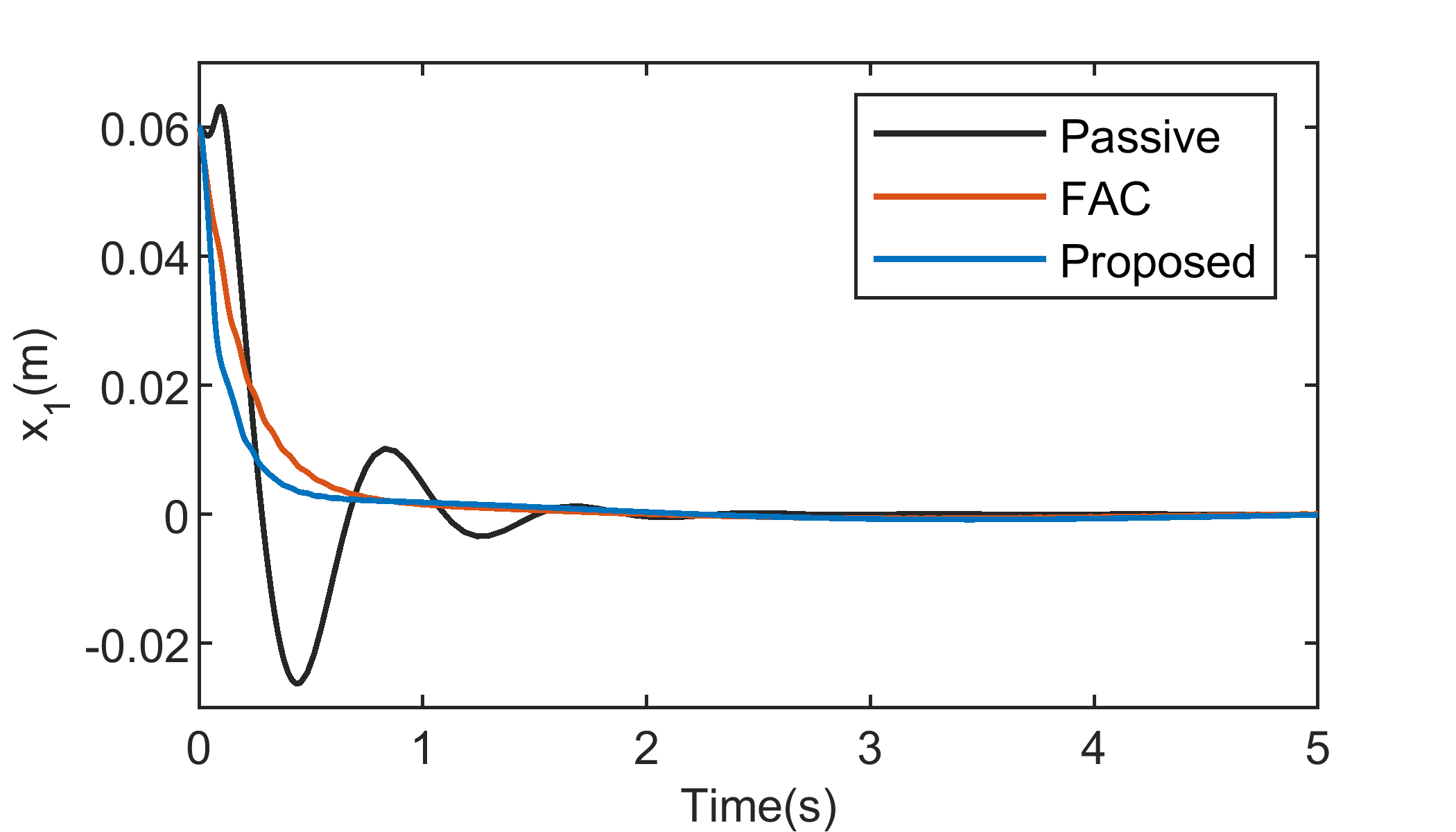}}
	\caption{Suspension system's response $x_1$ in bump road}
	\label{fig_71}
\end{figure}

\begin{figure}[!t]
	\centerline{\includegraphics[width=\columnwidth]{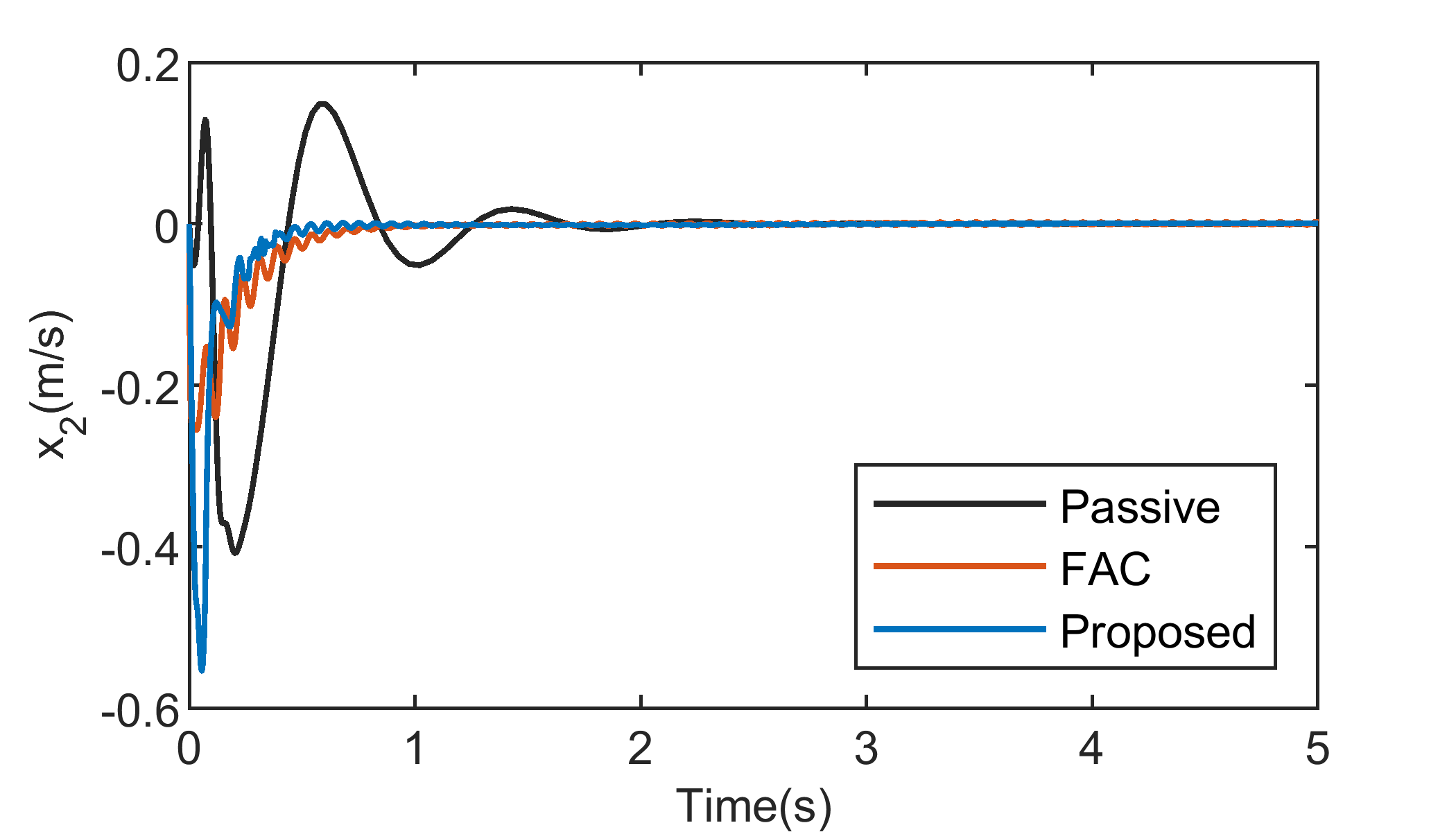}}
	\caption{Suspension system's response $x_2$ in bump road}
	\label{fig_72}
\end{figure}

\begin{figure}[!t]
	\centerline{\includegraphics[width=\columnwidth]{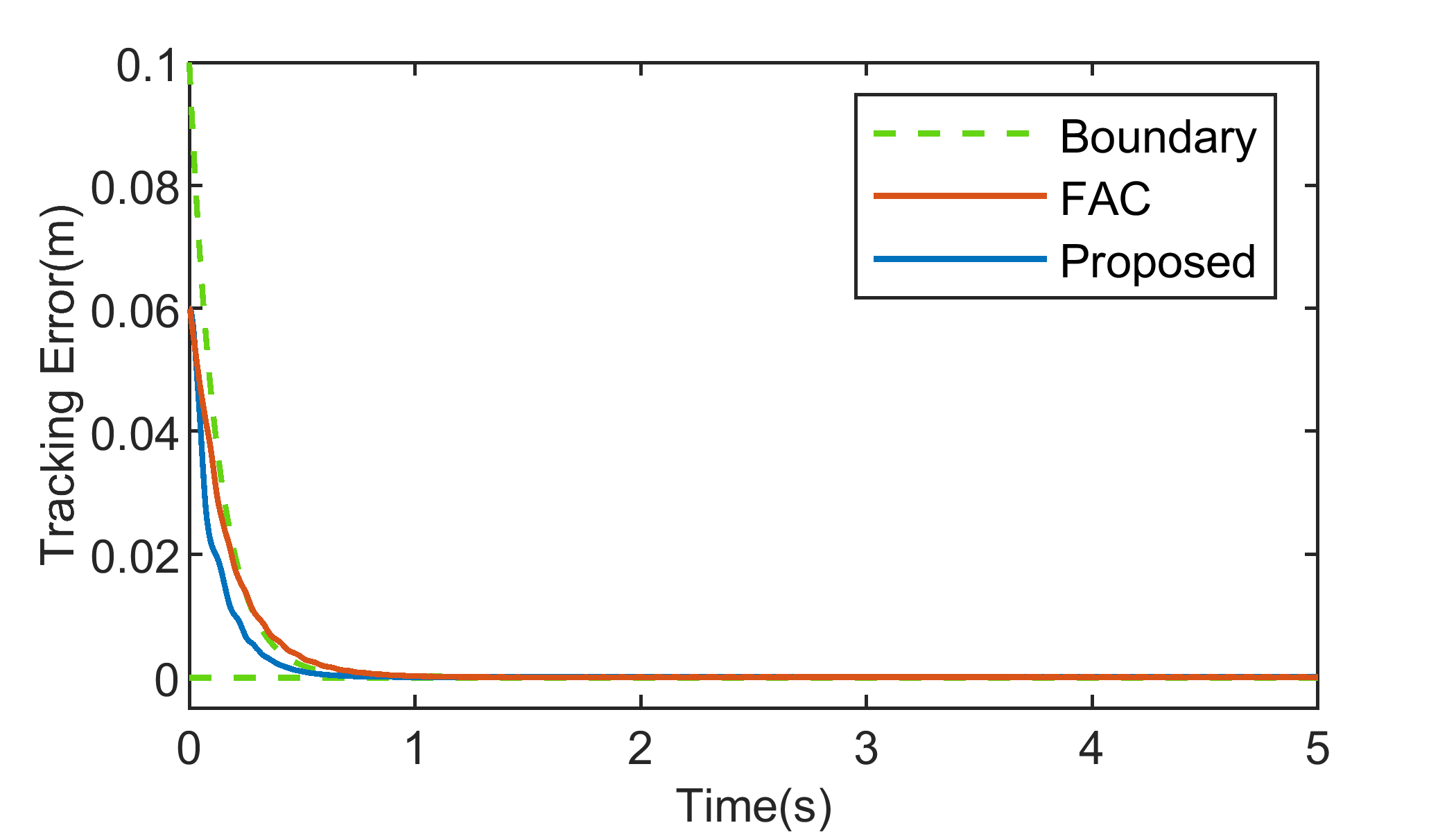}}
	\caption{Comparison of convergence performance between FAC and proposed controller in bump road}
	\label{fig_7}
\end{figure}

The responses of the suspension system to bump excitation under the control of the FAC and the proposed method are depicted in Fig \ref{fig_71} and Fig \ref{fig_72}, where the response of the passive suspension system is to demonstrate the control's effect more clearly. The absolute displacement $x_1$ is shown in Fig \ref{fig_71} and the corresponding velocity $x_2$ in Fig \ref{fig_72}. It can be concluded from Fig \ref{fig_71} that both control methods are effective to stabilize the suspension system to a small residual set as soon as possible and eliminate the fluctuation when bump excitation occurs. Moreover, the proposed method has a faster response than FAC, as shown in Fig \ref{fig_71}, and the velocity $x_2$ in Fig \ref{fig_72} also proves the conclusion in another way. The tracking error displayed in Fig \ref{fig_7} further indicates the faster convergence of the proposed method. In Fig \ref{fig_7}, the green dotted line is the tracking error convergence boundary preset by the prescribed performance function. The tracking error under the proposed controller is within the convergence boundary for the entire time, while a part of the  tracking error under the FAC does not, which indicates that the proposed controller can confine the tracking error in the prescribed range. Moreover, it can be seen from Fig \ref{fig_7} that the convergence rate of the system signal controlled by the proposed method is greater than that of FAC. This merit can be further demonstrated in sinusoidal road subsequently.

\subsection{Sinusoidal Road}
The sinusoidal excitation is expressed as ${z_r} = 0.025\sin (2\pi t)$. The initial states of the system are the same as that in the bump road. The control parameters are set as: ${k_1} = 3$, ${k_2} = 0.033$, ${k_3} = 4.167$ and tanh function parameters are ${\delta _{L1}} = {\delta _{U1}} = {\delta _{L2}} = {\delta _{U2}} = {\delta _{L3}} = {\delta _{U3}} = 1$. Prescribed performance functions are chosen as 
$${\rho _{U1}} = (0.1 - 0.0001){e^{ - 7t}} + 0.0001, {\rho _{L1}} = 0.0001$$
$${\rho _2} = (1 - 0.01){e^{ - 5t}} + 0.01$$
$${\rho _3} = (12.5 - 0.5){e^{ - 1t}} + 0.5$$

\begin{figure}[!t]
	\centerline{\includegraphics[width=\columnwidth]{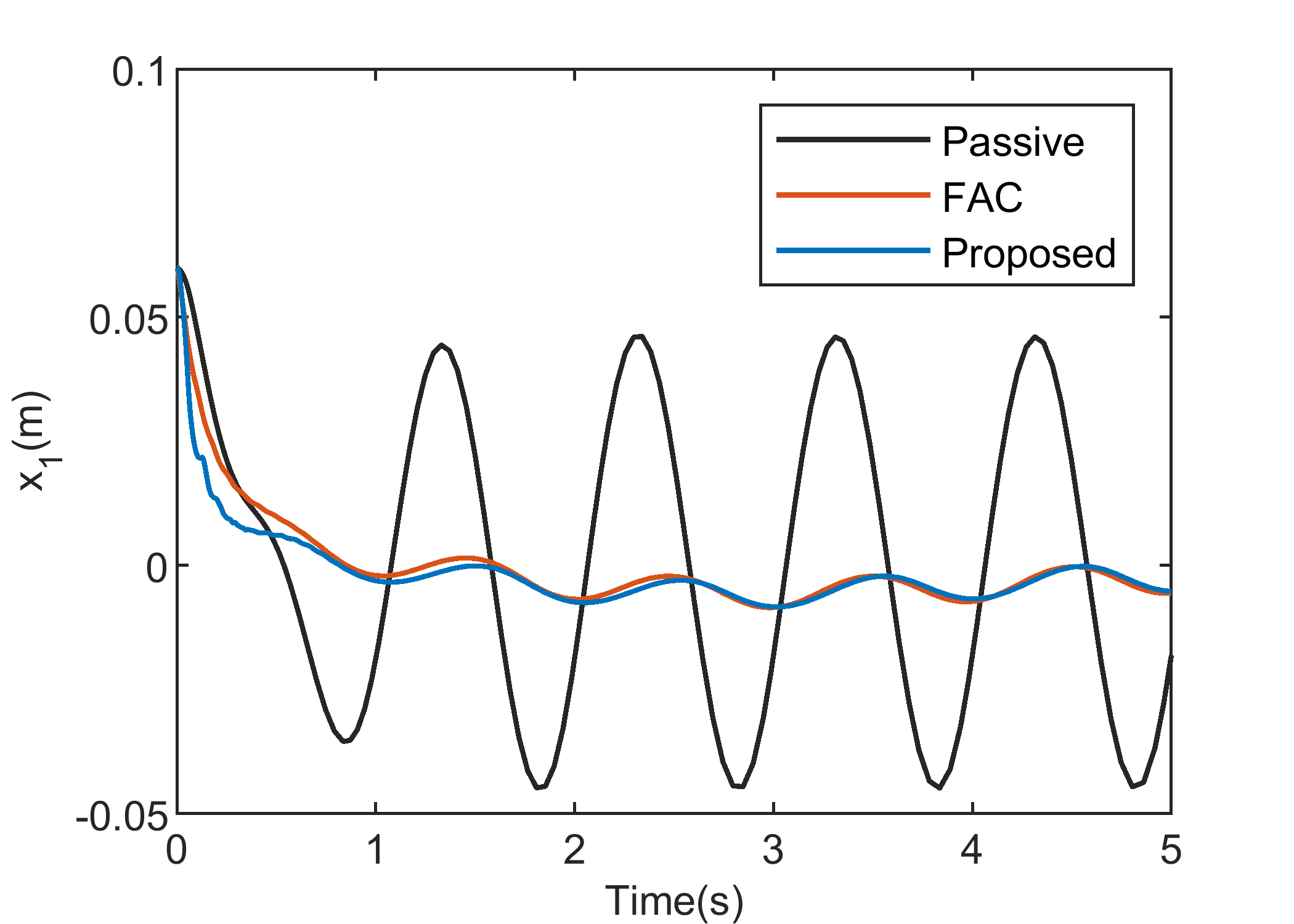}}
	\caption{Suspension system's response $x_1$ in sinusoidal road}
	\label{fig_81}
\end{figure}

\begin{figure}[!t]
	\centerline{\includegraphics[width=\columnwidth]{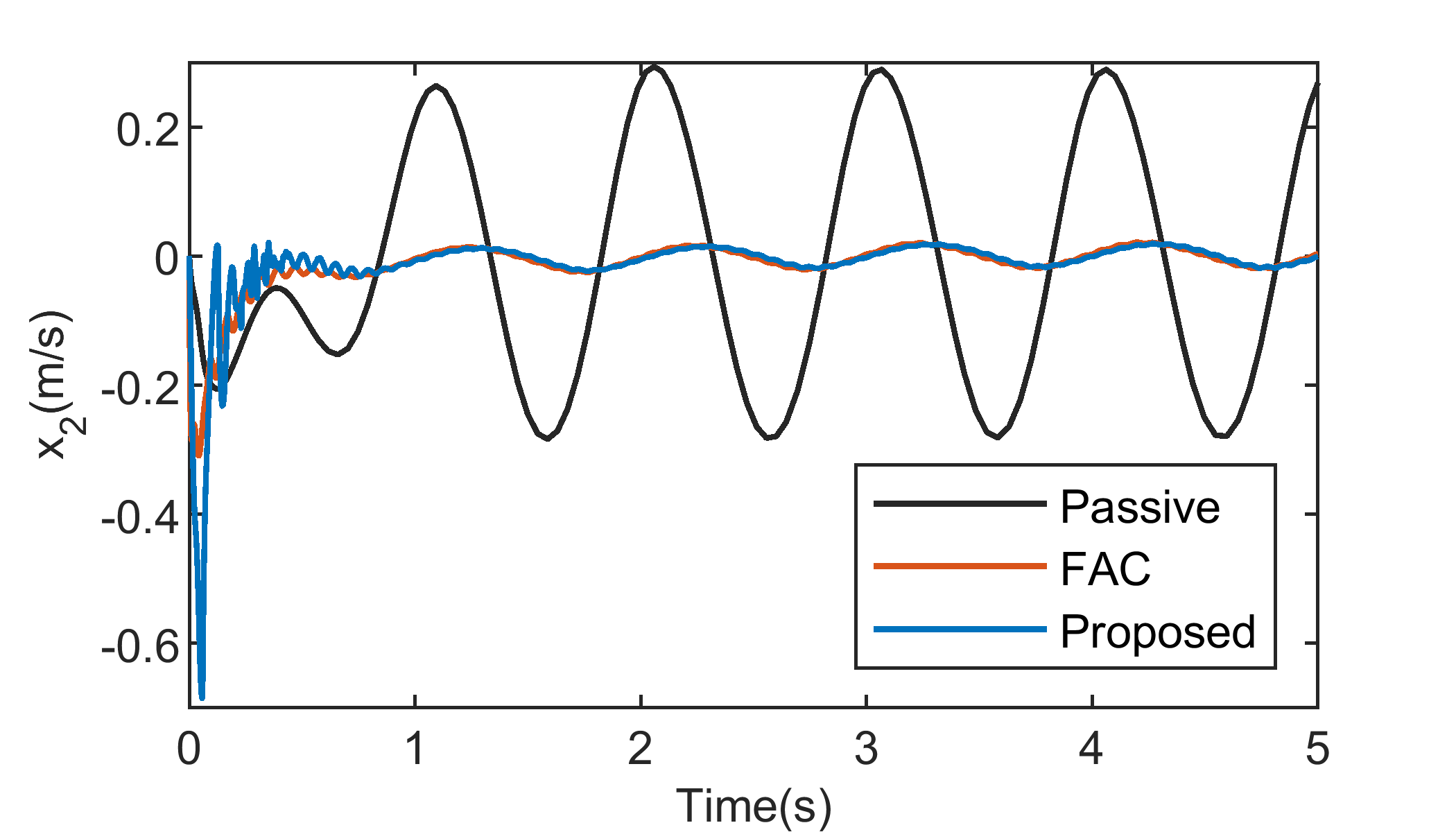}}
	\caption{Suspension system's response $x_2$ in sinusoidal road}
	\label{fig_82}
\end{figure}

\begin{figure}[!t]
	\centerline{\includegraphics[width=\columnwidth]{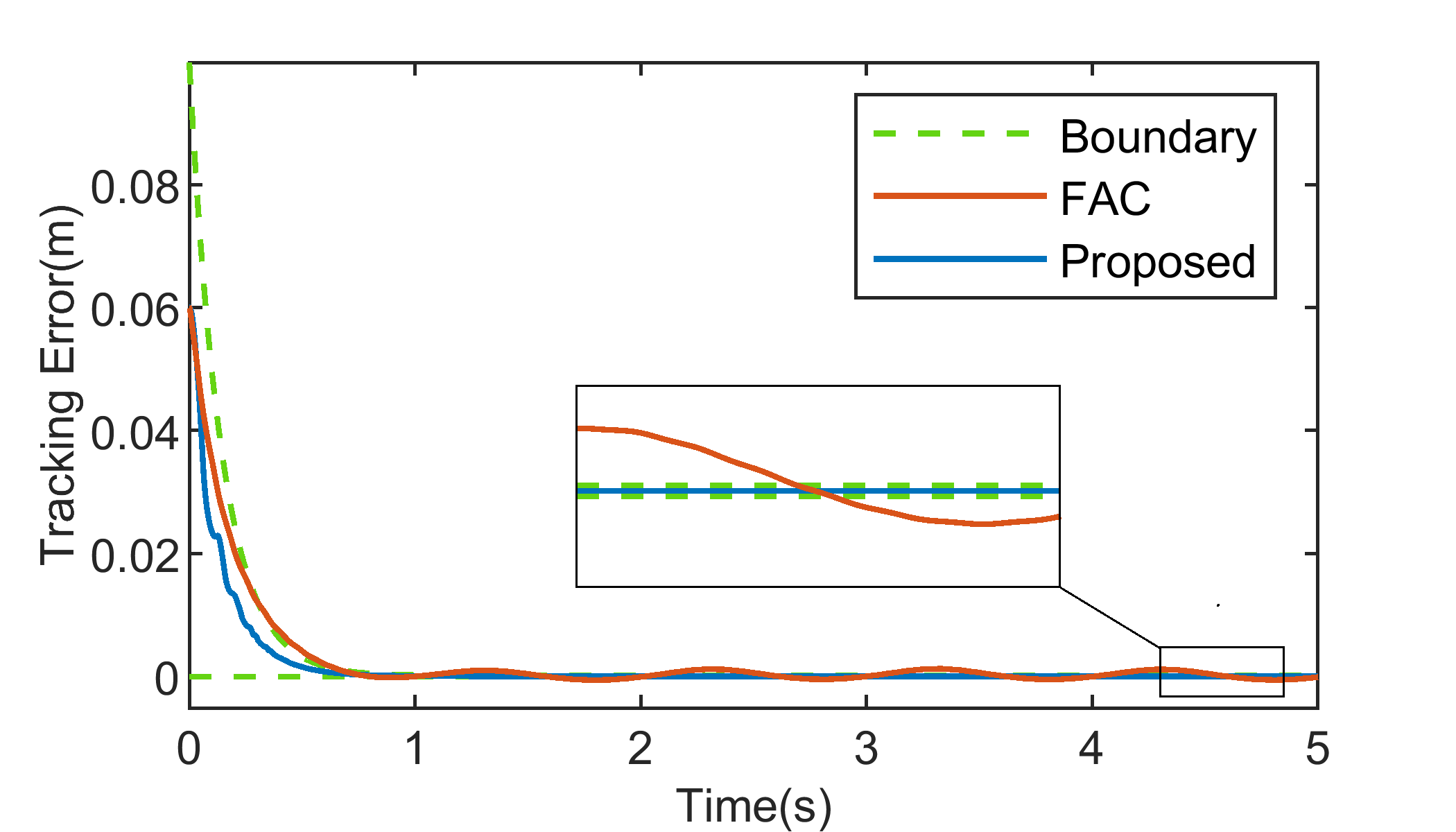}}
	\caption{Comparison of convergence performance between FAC and proposed controller in sinusoidal road}
	\label{fig_8}
\end{figure}

The responses of the passive suspension and the suspension under the control of  FAC and the proposed controller to sinusoidal excitation are described, and the absolute displacement  $x_1$ is displayed in Fig \ref{fig_81} and the corresponding velocity $x_2$ in Fig \ref{fig_82}. The results show the effectiveness of both two controllers in stabilizing the system and reducing fluctuation caused by sinusoidal excitation and the faster response of the proposed method can be seen when comparing the results in Fig \ref{fig_81} and Fig \ref{fig_82}. The further validation is in Fig \ref{fig_8}, which shows the tracking error of different controllers. The green dotted line has the same definition as mentioned in the bump road case. From Fig \ref{fig_8}, it can also be deduced that the proposed controller has a greater convergence rate than FAC. Moreover, it should be noted that the proposed method has a more precise tracking error than FAC, since the tracking error controlled by FAC violates the prescribed bound during the stable region while controlled by the proposed controller does not. It further indicates that the proposed controller possesses better convergence performance than FAC.

\section{Conclusion}
In this paper, a novel control scheme was proposed for a suspension system with uncertainty and unknown nonlinearity. The control scheme consisted two parts: one was the bioinspired model, and another one was the approximation-free control scheme. It has been proved by Lyapunov Theorem that the suspension system could be stabilized by the proposed control scheme. Main conclusions can be drawn from this work, as follows: Firstly, the significant computational time was reduced and the better vibration suppression performance was obtained without compromising ride safety, simultaneously. The proposed controller had a similar recursion scheme as backstepping control, but without adaptive law. Furthermore, the approximation functions (FLS or NNs) were avoided. Then it resulted in a simpler structure and less computational time. The nonlinearity inside the bioinspired model benefited in vibration suppression, then the ideal trajectory was produced and assisted the controller to suppress excess vibration. Moreover, a comparison simulation of time consumption and acceleration with or without disturbance and changeable passengers was conducted to justify the conclusion above.The results indicated the better vibration suppression performance, robustness, and simple structure of the proposed controller. Secondly, the assistance of the PPF involved in the proposed controller achieved better transient performancee. The exponentially decreased characteristic of  PPF promoted the convergence rate and arrived at a more precise tracking error. The theoretical comparison and analysis among these controllers with and without PPF  have been investigated, which confirmed the second conclusion. Moreover, comparison simulation carried in bump and sinusoidal road further verified the conclusion. The results illustrated the faster convergence and more precise response of proposed control scheme. In addition, this study process considered the safety and comfort requirements for the suspension system. It should be noted that the singularity of the approximation-free control remains still and many control parameters need to be determined synthetically by trial and error. Future work will integrate other control methods to solve the singularity problem and utilize optimization methods to design control parameters.

\appendices

\section*{Acknowledgment}

This work was supported by the National Natural Science Foundation of China (No. 11832009). The authors would like to express their gratitude to EditSprings (https://www.editsprings.com/) for the expert linguistic services provided.

\section{Bioinspired model's force analysis}
\label{appendix 1}
The dynamic analysis and the proof of Lemma1 are given here. As Fig \ref{fig_3} shown, the main force loads in joint 2. The force condition is analyzed in joint 1, joint 2 and joint 3, respectively. Then, we have the following equations satisfied.

\begin{equation}
	\label{eq59}
	\begin{split}
		f =& {k_v}{y_r} + \frac{{{k_h}}}{2}\left( {{L_1}\cos {\theta _1} + {L_2}\cos {\theta _2}} \right.\\
		&\left. { - \sqrt {L_1^2 - {v^2}({y_r})}  - \sqrt {L_2^2 - {v^2}({y_r})} } \right) \cdot \\
		&\left( {\frac{{{v^2}({y_r})}}{{\sqrt {L_1^2 - {v^2}({y_r})} }} + \frac{{{v^2}({y_r})}}{{\sqrt {L_2^2 - {v^2}({y_r})} }}} \right)
	\end{split}\tag{A1}
\end{equation}
where $v({y_r}) = {L_1}\sin {\theta _1} + {{{y_r}} \mathord{\left/
		{\vphantom {{{y_r}} 2}} \right.
		\kern-\nulldelimiterspace} 2}$. Then, according to the Lagrange principle, one can deduce the state-space equation \eqref{eq8} for the bioinspired model by following kinetic energy \eqref{eq60} and potential energy \eqref{eq61} and Hamilton function \eqref{eq62}
\begin{align}
	\label{eq60}\tag{A2}
	&T = \frac{1}{2}M{\dot y^2}\\
	\label{eq61}\tag{A3}
	&V = \frac{1}{2}{k_h}{x^2} + \frac{1}{2}{k_v}{({y_r} - \Delta y)^2} + Mgy\\
	\label{eq62}\tag{A4}
	&\frac{d}{{dt}}\left( {\frac{{\partial L}}{{\partial \dot y}}} \right) - \frac{{\partial L}}{{\partial y}} =  - {\mu _1}{\dot y_r} - {\mu _2}{n_x}\dot \varphi \frac{{\partial \varphi }}{{\partial y}}
\end{align}
where $\Delta y$ is the vertical spring compressive deflection and  $x$ satisfies that $x = {x_1} + {x_2}$. In Hamilton function, $L$ and $\varphi $ satisfy $L = T - V$ and $\varphi  = {\varphi _1} + {\varphi _2}$, respectively. ${n_x}$ is the number of joints. ${\mu _1}$ and ${\mu _2}$ are the air damping coefficient and rotational friction coefficient, respectively. 

\bibliographystyle{ieeetr}
\bibliography{reference0409}
\end{document}